\documentclass{aastex631}
\usepackage{tablefootnote}
\usepackage{}

\newcommand{\kms}{{\rm km\,s^{-1}}}

\newcommand{\msun}{{M}_\odot}

\begin{document}

\title{Ring Gap Structure around Class I Protostar WL~17}


\author[0000-0001-6580-6038]{Ayumu Shoshi}
\affiliation{Department of Earth and Planetary Sciences, Graduate School of Science, Kyushu University,\\744 Motooka, Nishi-ku, Fukuoka 819-0395, Japan}
\author[0000-0002-8217-7509]{Naoto Harada}
\affiliation{Department of Earth and Planetary Sciences, Graduate School of Science, Kyushu University,\\744 Motooka, Nishi-ku, Fukuoka 819-0395, Japan}
\author[0000-0002-2062-1600]{Kazuki Tokuda}
\affiliation{Department of Earth and Planetary Sciences, Faculty of Science, Kyushu University,\\744 Motooka, Nishi-ku, Fukuoka 819-0395, Japan}
\affiliation{National Astronomical Observatory of Japan, 2-21-1 Osawa, Mitaka, Tokyo 181-8588, Japan}
\author[0000-0002-9424-1192]{Yoshihiro Kawasaki}
\affiliation{Department of Earth and Planetary Sciences, Graduate School of Science, Kyushu University,\\744 Motooka, Nishi-ku, Fukuoka 819-0395, Japan}
\author{Hayao Yamasaki}
\affiliation{Department of Earth and Planetary Sciences, Graduate School of Science, Kyushu University,\\744 Motooka, Nishi-ku, Fukuoka 819-0395, Japan}
\author[0000-0001-5817-6250]{Asako Sato}
\affiliation{Department of Earth and Planetary Sciences, Graduate School of Science, Kyushu University,\\744 Motooka, Nishi-ku, Fukuoka 819-0395, Japan}
\author[0000-0002-7951-1641]{Mitsuki Omura}
\affiliation{Department of Earth and Planetary Sciences, Graduate School of Science, Kyushu University,\\744 Motooka, Nishi-ku, Fukuoka 819-0395, Japan}
\author[0000-0002-8185-9882]{Masayuki Yamaguchi}
\affiliation{Academia Sinica Institute of Astronomy and Astrophysics, 11F of ASMA Building, No.1, Sec. 4, Roosevelt Rd, Taipei 106, Taiwan}
\author[0000-0002-1411-5410]{Kengo Tachihara}
\affiliation{Department of Physics, Nagoya University, Furo-cho, Chikusa-ku, Nagoya 464-8601, Japan}
\author[0000-0002-0963-0872]{Masahiro N. Machida}
\affiliation{Department of Earth and Planetary Sciences, Faculty of Science, Kyushu University,\\744 Motooka, Nishi-ku, Fukuoka 819-0395, Japan}

\begin{abstract}
WL~17 is a Class I object and was considered to have a ring--hole structure.
We analyzed the structure around WL~17 to investigate the detailed properties of WL~17.
We used ALMA archival data, which have a higher angular resolution than previous observations.
We investigated the WL~17 system with the 1.3\,mm dust continuum and $^{12}$CO and C$^{18}$O ($J$ = 2--1) line emissions.
The dust continuum emission showed a clear ring structure with inner and outer edges of $\sim$\,11 and $\sim$\,21\,au, respectively.
In addition, we detected an inner disk of $<$\,5\,au radius enclosing the central star within the ring, the first observation of this structure.
Thus, WL~17 has a ring--gap structure, not a ring--hole structure.
We did not detect any marked emission in either the gap or inner disk, indicating that there is no sign of a planet, circumplanetary disk, or binary companion.
We identified the base of both blue-shifted and red-shifted outflows based on the $^{12}$CO emission, which is clearly associated with the disk around WL~17.
The outflow mass ejection rate is $\sim$3.6$\times10^{-7}\,M_{\odot} \,{\rm yr}^{-1}$ and the dynamical timescale is as short as $\sim10^4$\,yr.
The C$^{18}$O emission showed that an inhomogeneous infalling envelope, which can induce episodic mass accretion, is distributed in the region within $\sim1000$\,au from the central protostar.
With these new findings, we can constrain the planet formation and dust growth scenarios in the accretion phase of star formation.
\end{abstract}


\section{Introduction} \label{sec:intro}
Stars (or protostars) form in molecular cloud cores, and planets form in the course of star formation. 
The typical mass and density of molecular cloud cores are the order of $\sim1\,M_{\odot}$ with a H$_2$ volume density of 10$^{4}$--10$^{6}$\,cm$^{-3}$ \citep[e.g.][]{Onishi_2002,Tachihara_2002,Andre_2014,Tokuda_2020}. 
Protostars form  in gravitationally collapsing cloud cores  \citep{Larson_1696}. 
Since the mass of a protostar at its birth is $\sim 10^{-3}\,M_{\odot}$ \citep{Masunaga_2000}, a remnant of the collapsing cloud core (or infalling envelope) with a mass of $\sim1\,M_{\odot}$ remains around the newborn star \citep{bate1998,Machida_2011}. 
The protostar then grows by accretion from the infalling envelope \citep[for details, see review by][]{Tsukamoto_2022}. 
The mass accretion from the envelope lasts for about $\sim10^5$\,yr \citep{Enoch_2009} during which gas is supplied from the infalling envelope to the protostar through the circumstellar disk. 
The circumstellar disk also grows in the mass accretion phase \citep[e.g.,][]{Tomida_2017}. 
Part of the envelope gas accretes onto the circumstellar disk and protostar, while the remainder is ejected by the protostellar outflow \citep{Matzner_2000,Machida_2013}. 
After the infalling envelope is depleted, the circumstellar disk or protoplanetary disk gradually dissipates. 
Finally, the central object (protostar or pre-main sequence star) contracts on the Kelvin--Helmholtz timescale and hydrogen burning occurs, creating a main sequence star.

It has been considered that planet formation begins in the protoplanetary disk after the mass accretion phase ends \citep{Hayashi_1981,Hayashi_1985}. 
In other words, planet formation occurs in an isolated disk and the mass supply from the infalling envelope and mass ejection by the outflow are ignored.   
The first step of planet formation is dust growth \citep[see review by][]{Testi_2014}. 
Observations have confirmed sub-micron-sized dust grains in interstellar space and molecular cloud cores \citep{Lada_2003,Williams_2011}. 
The collisional growth of such dust grains is possible in high-density regions \citep[e.g.,][]{kawasaki_2022,kawasaki_2023}. 
Past studies have found that it is difficult for dust grains to grow in gravitationally collapsing cloud cores \citep{Ormel_2009,Ormel_2011,Hirashita_2013}. 
Note that some observations have also shown possible evidence of micron- and millimeter-sized dust in core and envelope scales \citep[e.g.,][]{Miotello_2014,Lefevre_2016}.
Recent theoretical studies indicate that the growth of dust grains with the size larger than millimeters is possible in circumstellar disks during the main accretion stage \citep{Vorobyov_2018,Tsukamoto_2021,Ohashi_2021,koga_2023}. 
It is thus important to identify when planet formation or dust growth begins in the star formation process. 

Recent observations imply  that planet formation begins before  the end of the accretion stage or Class 0/I stages \citep{Tychoniec_2020}. 
In the following, we call a disk around a Class II object a protoplanetary disk and a disk around a Class 0 or I object a circumstellar disk. 
\citet{Tychoniec_2020} have shown that there is insufficient dust to produce several planets in protoplanetary disks after the accretion phase. 
In addition, various structures such as rings, gaps, holes, and spiral designs have been confirmed in many protoplanetary disks \citep[][]{Andrews_2018}. 
These observations also imply that planet formation or dust growth may begin earlier than the Class II stage.
Note that rings, gaps, holes, and spirals do not necessarily imply the existence of planets because a gravitationally unstable disk can form such features.
It is difficult to observe signs of planet formation in circumstellar disks around Class 0 objects because such disks are still deeply embedded in dense gas envelopes.
In addition, the disks around Class 0 objects tend to be massive and optically thick, making it difficult to detect forming planets.
However, it is important to observe disks in the early stage of star formation to catch the first signs of planet formation.

Circumstellar disks around Class I objects are plausible sites for confirming the signs of planet formation because we can observe disks with less massive infalling envelopes \citep{Marana_2018,Williams_2019,Sanchis_2020,Mulders_2020}. 
In addition, a sufficient amount of dust remains in circumstellar disks around Class I objects \citep{Tychoniec_2020}. 
Recent observations have shown a ring--gap or ring--hole structure in circumstellar disks around Class 0/I objects \citep{Sheehan_2020,Segura_cox_2020}.

Our target WL~17 is a protostar in the $\rho$ Ophiuchus L1688 molecular cloud located 137\,pc from the Sun \citep[e.g.,][]{Ortiz-Leon_2017,Ortiz-Leon_2018}. 
\citet{Enoch_2009} and \citet{van_Kempen_2009} have shown that WL~17  is a Class I object.
They estimated that the age of the protostar WL~17 is younger than $\sim5\times$10$^5$ yr.
\citet{Sheehan_2017} also verified WL~17  as a Class I protostar by spectral energy distribution (SED) model fitting.
They found that WL~17 has a ring--hole structure based on high angular resolution ALMA Band 3 observations, where the radius of the hole is $\sim10$\,au and the outer radius of the ring is $\sim20$\,au. 
Thus,  the Class I object WL~17 should be a promising target for clarifying when planet formation and dust growth begin.  

Although past observations have indicated that WL~17 is a Class I object, it is not usual for a Class I object to have a ring structure.
In addition, some suspicious characteristics of the WL~17 system make it difficult to identify WL~17 as a Class I object.
The bolometric luminosity of WL~17 is about 0.6\,$L_{\odot}$\citep{Sheehan_2017}, which is somewhat lower than typical Class I objects. 
Note that it has been well known since the 1990s that some Class I objects exhibit  luminosities lower than $< 1L_{\odot}$ \citep{Kenyon_1990}.

In addition,  N$_2$H$^+$, DCO$^+$, and C$^{18}$O line emissions, indicators of the system youth, were not detected around WL~17 in past studies \citep{Loren_1990,Andre_2007,van_Kempen_2009}.  
Then, using a single-dish telescope, \citet{van_der_Marel_2013} showed a broad $^{12}$CO ($J$ = 3--2) line emission, implying an outflow associated with WL~17.
However, it is difficult to make conclusions about the outflow driven by the WL~17 system because of the insufficient spatial resolution.
Thus, the presence of an outflow has yet to be confidentially confirmed around WL~17.

Class I objects are in the accretion phase, meaning that the infalling envelope encloses the central object. 
The typical accretion luminosity is expected to be $\lesssim 1$--$10$\,$L_{\odot}$ \citep{Scott_1995}. 
In addition, the outflow is a useful indicator of gas accretion because the release of the gravitational energy of the accreting matter drives the outflow \citep{Pudritz_1986,Tomisaka_2002}.  
On the other hand,   ring and hole structures are considered to appear in the later Class II evolutionary stage.  
Thus, more detailed observations of the WL~17 system would constrain the formation of substructures in the disk to the early phase of star and planet formation.

The aim of this study is to clarify the evolutionary stage and growth timescale of the substructure of WL~17 using two sets of high-resolution ALMA archival data.
The data contain continuum and $^{12}$CO and C$^{18}$O ($J$ = 2--1) line emissions. 
Details of the data are presented in \S\ref{sec:obs}. 
In \S\ref{sec:results},  we present the results of the continuum and $^{12}$CO line emissions and estimate the physical quantities of the WL~17 system, such as the gas mass and outflow mass ejection rate.
In \S\ref{sec:discussion}, we confirm the details of the remaining envelope gas around WL~17 with C$^{18}$O line observation and Herschel's H$_2$ column density map. 
In addition, we discuss the possible process of substructure formation around WL~17 and suggest plausible scenarios to explain the WL~17 system in \S\ref{sec:discussion}. 
We summarize our results in \S\ref{sec:summary}. 

\section{Observation Data and Imaging} \label{sec:obs}
\subsection{High-resolution Continuum Observations}
We used ALMA archival data (Project 2019.1.00458.S, PI Patrick Sheehan) to obtain high-resolution continuum images.
The observations were carried out on 2019 September 29 using the ALMA main array in its C43-9/10 configuration.
There were two continuum spectral windows with central frequencies of 218\,GHz and 232\,GHz in Band 6.
Both had a bandwidth of 1.875\,GHz.

The data were calibrated with the Common Astronomy Software Application \citep[CASA;][]{McMullin_2007} version 6.2.1. 
We used only the \texttt{tclean} task in our imaging process.
We applied Briggs weighting with a robustness parameter of 0.5 and an image grid of 0\farcs0037.
We performed self-calibrations and could not achieve any improvement because of the low signal-to-noise-ratio ($<$100).
The resultant synthesized beam size was $0\farcs037 \times 0\farcs031$ ($5.1 \times 4.2$\,au) with a position angle (PA) of $-65.9^\circ$ at the distance of WL~17 \citep[137\,pc;][]{Sheehan_2017}, which is the highest resolution ever achieved for this source.
The sensitivity of the continuum achieved was 18\,$\mu$Jy\,beam$^{-1}$.

\subsection{Molecular Line Observations}
We used further ALMA archival data (Project 2019.1.01792.S, PI Diego Mardones) to investigate the gas structure around the protostellar disk.
The observations were carried out on 2019 November 21 using the ALMA main array in its C-7 configuration.
We used three spectral windows targeting the continuum, $^{12}$CO\,($J$ = 2--1), and C$^{18}$O\,($J$ = 2--1).
The continuum spectral window had a central frequency of 232\,GHz and a bandwidth of 1.875\,GHz.
The $^{12}$CO and  C$^{18}$O line spectral windows had central frequencies of 230 and 219\,GHz, respectively.
The bandwidth of both windows was 11.7\,MHz.

In the imaging process, we also used the CASA task \texttt{tclean} and applied Briggs weighting with a robustness parameter of 0.5, an image grid of 0\farcs140, and a velocity resolution of 0.1\,km\,s$^{-1}$.
In addition, the continuum was subtracted from the $^{12}$CO and C$^{18}$O line data using the CASA task \texttt{imcontsub}.
The resultant synthesized beam size was $1\farcs21 \times 0\farcs93$ ($166 \times 127$\,{au}) with a PA of $-85.5^\circ$.
We achieved a sensitivity  of 0.33\,mJy\,beam$\mathbf{^{-1}}$ for the continuum and 0.35\,K per 0.1\,km\,s$^{-1}$ for the CO lines.

\begin{figure*}[t]
    \centering
    \includegraphics[width=0.9\linewidth]{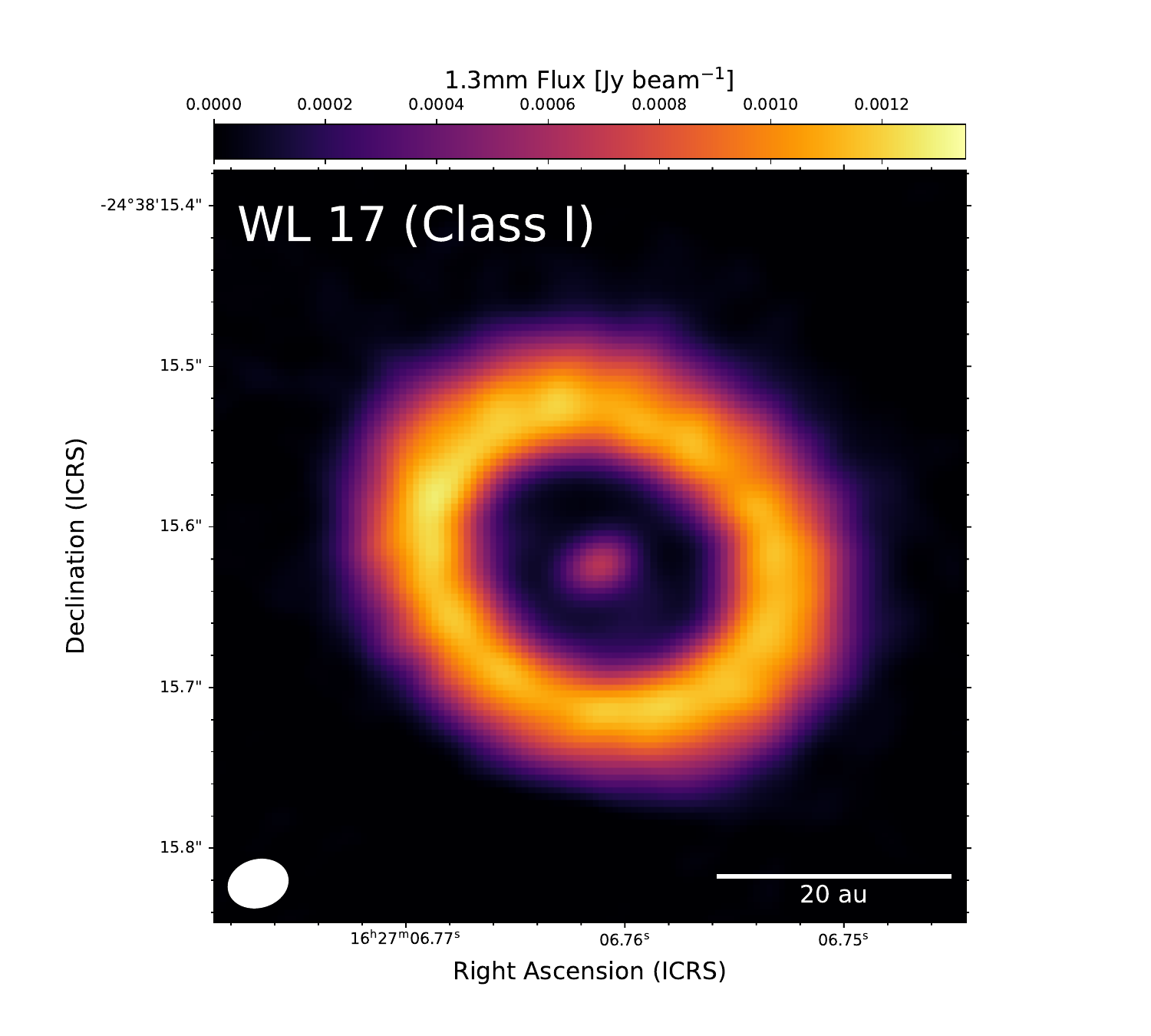}
    \caption{ALMA 1.3\,mm continuum image of WL~17. 
    The white ellipse in the lower left corner denotes the synthesized beam; 0\farcs037 $\times$ 0\farcs031 (5.1 $\times$ 4.2\,au) with a PA of $-$65.9$^{\circ}$.}
 \label{fig:cont}
\end{figure*}

\begin{figure*}[t]
    \centering
    \includegraphics[width=\linewidth]{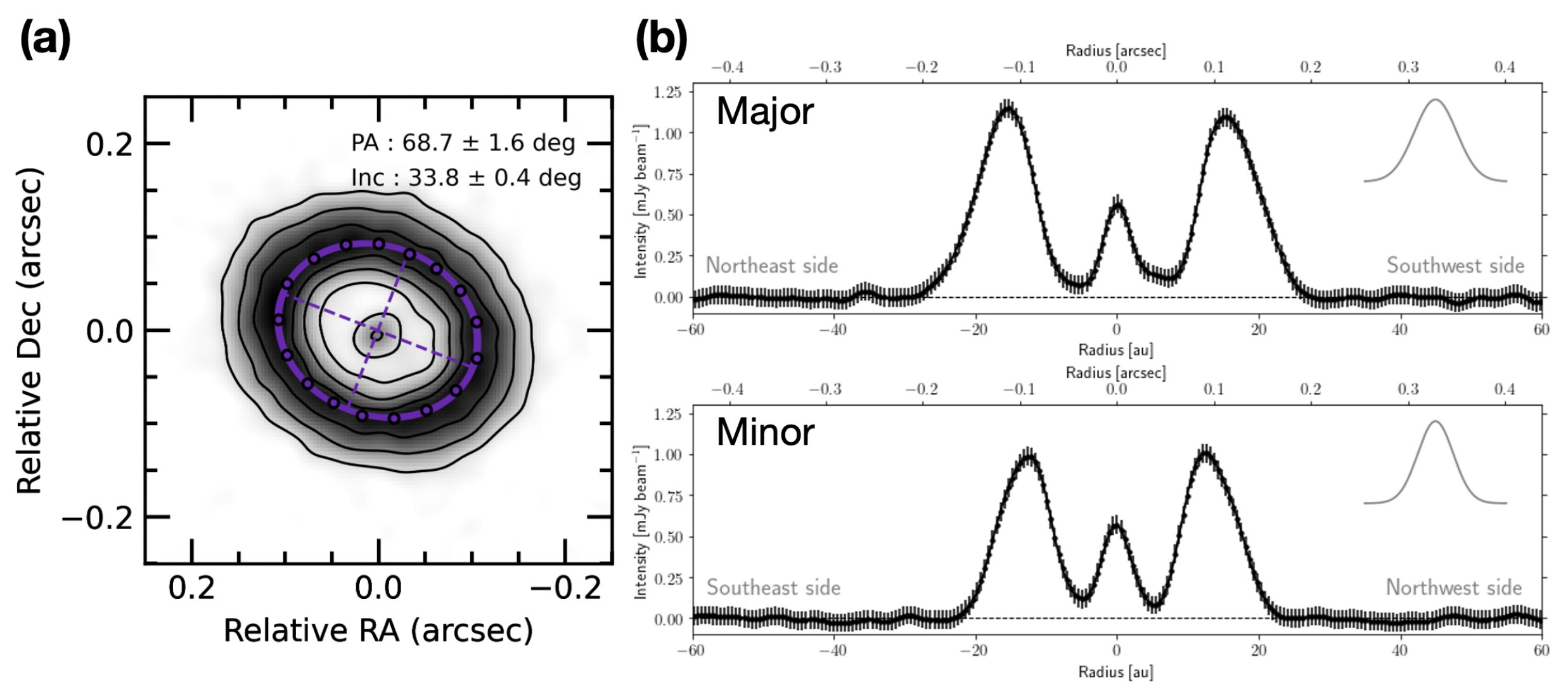}
    \caption{(a) Best-fit ellipse (purple dashed-dotted curve) for the disk (or ring) overlaid on the 1.3\,mm continuum image with contours 10, 30, 50 and 70\,$\sigma$ (where $1\sigma$=18$\mu$\,Jy\,beam$^{-1}$).
    The major and minor axes are represented by the purple broken lines. 
    The inclination (Inc) and position (PA) angles are described in the upper right corner. 
    (b) Intensity profiles along the major (top) and minor (bottom) axes.
    The error bars correspond to 3$\sigma$. 
    The grey curve in the top right of each panel shows the profile for the synthesized beam.}
 \label{fig:intpro}
\end{figure*}

\begin{figure*}[ht]
    \centering
    \includegraphics[width=0.6\linewidth]{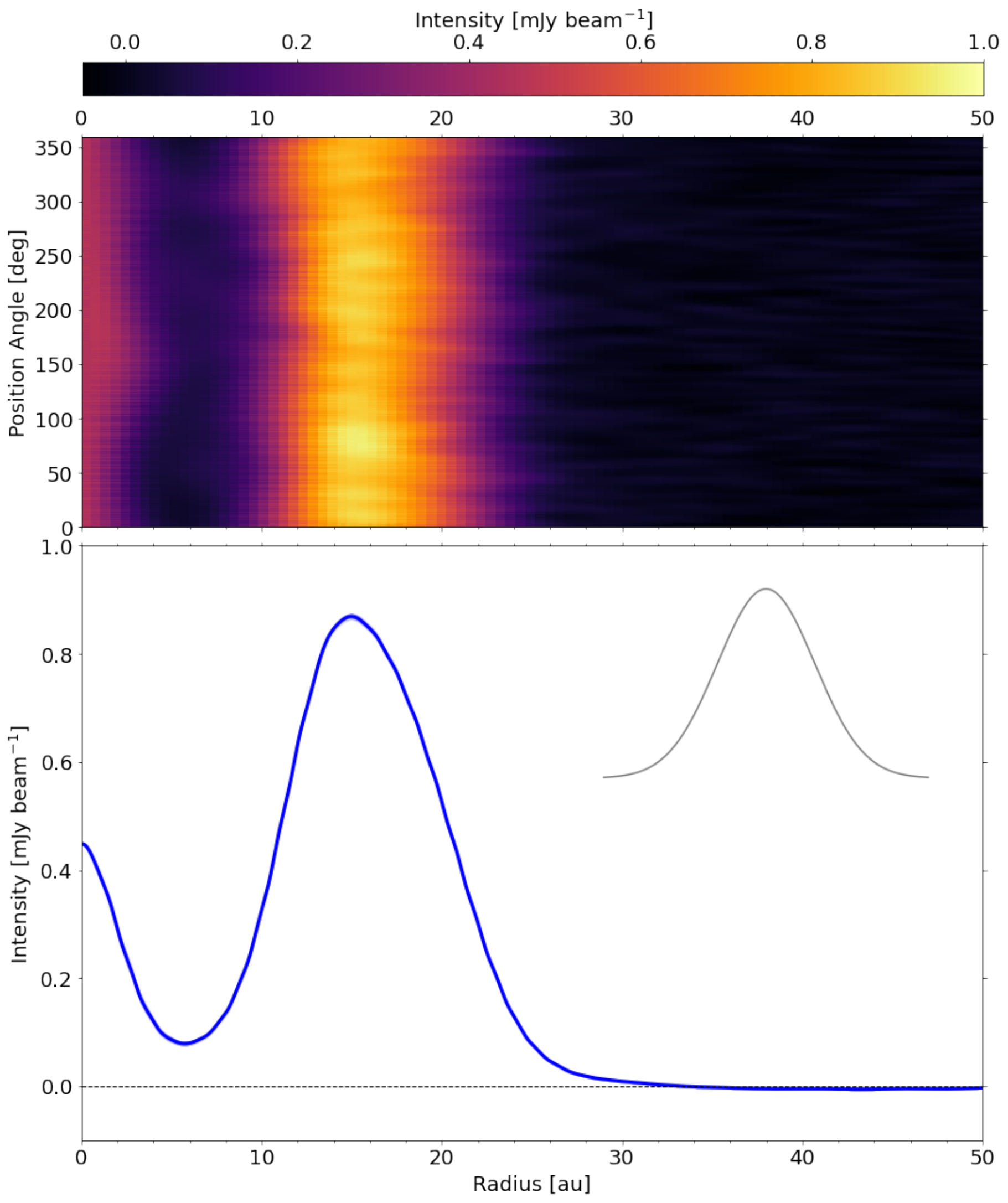}
    \caption{(Top) 1.3\,mm continuum emission map projected in polar coordinates (or on the radius and position angle plane, see \citealt{Yamaguchi_2021}).
    (Bottom) Radial intensity profile averaged over azimuthal direction.
    The profile is linearly interpolated onto radial grid points spaced by 0.1\,au using \texttt{interpolate.interp1d} in the \texttt{Scipy} module.
    The standard deviation is 2.9$\times$10$^{-3}$\,mJy\,beam$^{-1}$.
    The grey curve in the top right is the profile for the synthesized beam.
    }
 \label{fig:azintpro}
\end{figure*}

\section{Results}\label{sec:results}
\subsection{High-spatial Resolution Continuum Image} \label{sec:cont}

Figure~\ref{fig:cont} shows a 1.3\,mm (Band6; 225\,GHz) continuum image of WL~17 at an angular resolution of 0\farcs037 $\times$ 0\farcs031.
The geometric mean of the major and minor axes of the beam is 1.6 times smaller than that in the previous 3\,mm (Band 3; 97.5\,GHz) observations \citep{Sheehan_2017}.
We smoothed our 1.3\,mm image into a lower-resolution image with a beam size of 0\farcs06$\times$0\farcs05, which is the same as in \citet{Sheehan_2017}. Then, we obtained the sensitivity of $\sim$21\,$\mu$Jy\,beam$^{-1}$. 
Assuming a spectral index $\alpha$=2 \citep[or $\beta=0$;][]{Ilseing_2023}, our imaging sensitivity can be equivalently converted into $\sim$4\,$\mu$Jy\,beam$^{-1}$ at 3\,mm, which is nine times higher than that in the previous 3\,mm observation, 36\,$\mu$Jy beam$^{-1}$.
Although \cite{Sheehan_2017} mentioned the intensity enhancement of the central position, the contrast between the central position and surrounding is marginal.
Thus, they interpreted the inner structure as a hole with a radius of 13\,au. 
In our study, the improved sensitivity and angular resolution clearly illustrate the presence of an inner disk, which is expected to enclose the central protostar.
Thus, WL~17 has a ring--gap structure.

In the following, we refer to the inner dust disk and outer dust ring observed in the 1.3 mm continuum (Fig.~\ref{fig:cont}) as the inner disk and the outer ring, respectively. 
We also use the term $`$disk' to describe the disk structure around WL~17. 
While the gas component of the disk around WL~17 has not been clearly resolved in molecular line emissions, it is expected to be present around the Class I object. 
In this paper, we consider the whole disk to contain undetected gas and dust, which are distributed to a large radius ($\gg 20$\,au).
The inner disk and outer ring are part of the disk.
In addition, the outflow is considered to be driven by the disk or the whole disk region.

We fitted the outer ring with an ellipse and derived the inclination angle and position angle (PA) for the outer ring on the image shown in Figure~\ref{fig:cont}, using the method adopted in \citet{Yamaguchi_2021}. 
The fitted ellipse is shown in Figure~\ref{fig:intpro}(a). 
We describe the fitting procedure in the following. 
First, we determined the radial peak position of the outer ring every 1$^{\circ}$ in the position angle (or vertical) direction from the intensity map on polar coordinates.
Second, these collective peak positions were plotted  on the continuum intensity map (purple broken line in Fig.~\ref{fig:intpro}a). 
Using the major and minor axes of the ellipse (or the purple broken line), we estimated an inclination angle $i_{\rm ring}$ of 
33.8$^{\circ}$ and a PA of 68.7$^{\circ}$, as shown in Figure~\ref{fig:intpro}(a),
indicating that we observed the WL~17 system nearly face-on.

Figure~\ref{fig:intpro}(b) shows the intensity profiles in the major- and minor-axis directions.
The size of the outer ring is measured to be 56\,$\times$\,46\,au with PA = 68.7$^{\circ}$. 
Using the derived ellipse fitting parameters $i_{\rm disk}$ and PA, we deprojected the 1.3\,mm continuum image of the disk.
As shown in the top panel of Figure~\ref{fig:azintpro}, we created the intensity map on polar coordinates from the deprojected image.
The azimuthally averaged radial intensity profile is plotted in the bottom panel of Figure \ref{fig:azintpro}.
The uncertainty in the radial profile can be evaluated as the error of the mean at each radius, with a value of $2.9\times10^{-3}$\,mJy\,beam$^{-1}$.
We adopted the Gaussian fitting to the peaks of the inner disk and the outer ring shown in Figure~\ref{fig:azintpro} and estimated the full width at half maximum (FWHM) without the beam deconvolution.
The deconvolved size in the FWHM of the inner disk is 5.2\,au.
The size of the inner disk is comparable to the beam size, making it difficult to resolve its internal structure.
Thus, for the inner disk, we consider the estimated FWHM ($5.2$\,au) as the upper limit of the radius.
On the other hand, the peak radius of the outer ring is 15.7\,au and the FWHM is 8.8\,au, suggesting that the outer ring can be resolved spatially (the inner and outer edge radius of the outer ring are 11 and 21\,au, respectively).
Furthermore, we could not detect any strong thermal dust emission within the gap between the outer ring and inner disk in the range of $\sim5$--$10$\,au even in the high-spatial-resolution image (Fig.~\ref{fig:azintpro} top).
Figures~\ref{fig:intpro} and \ref{fig:azintpro} indicate that the intensity of thermal dust emission in the gap is significantly lower than that of the outer ring.
Thus, there is a very high-intensity contrast between the gap and the outer ring.

We measured the total flux emitted from the ring--gap structure of WL~17 system (Fig.~\ref{fig:cont}) considering the dust emission larger than 5$\sigma$ ($\sigma$ = 18\,$\mu$Jy beam$^{-1}$).
The total flux is $F_{\nu} = 4.37 \times$10$^{-2}$\,Jy, which is the sum of the inner disk (6.6$\times$10$^{-4}$\,Jy) and outer ring (4.3$\times$10$^{-2}$\,Jy). 
Thus, the flux from the outer ring is much greater than that from the inner disk. 
With the total flux $F_{\nu}$, we estimated the dust mass $M_{\rm dust}$ seen in Figure~\ref{fig:cont} to be
\begin{equation}
M_{\rm dust} = \frac{F_{\nu}\,d^2}{\kappa_{\nu}B _{\nu}(T_{\rm dust} )}, 
\end{equation}
where the dust is assumed to be optically thin, and a dust opacity at 1.3 mm of $\kappa_{\nu}=2$\,cm$^2$ g$^{-1}$ \citep{Beckwith_1990} and a dust temperature of $T_{\rm dust}=20$\,K were adopted. 
The total dust mass was estimated to be $M_{\rm dust}$=7.4$\times$10$^{-5}\,M_{\odot}$. 
Adopting a gas-to-dust mass ratio of 100, the outer ring mass of WL~17 is at least $M_{\rm gas,ring}\simeq 0.01 M_{\odot}$, which is comparable or smaller than the dust ring mass estimated in previous studies ($M_{\rm gas,ring}\simeq 0.04\,M_{\odot}$ for \citealt{Sheehan_2017}).  
Note that we only estimated the mass of the outer ring shown in Figure~\ref{fig:cont}. 
Thus,  $M_{\rm gas,ring}$ in our estimate should be the lower limit of the disk mass around WL~17 (see also \S\ref{sec:uncertainty}). 
The outer ring mass $M_{\rm gas,ring}$ around WL~17 estimated in this study is comparable to or slightly smaller  than the disks around Class I objects \citep{Eleonora_2023}. 

We compare the properties of the inner disk of the WL~17 system with those of other inner disks of objects associated with the ring--gap structure presented in \citet{Francis_2020}.
Note that the main targets of \citet{Francis_2020} are not circumstellar disks but protoplanetary disks.
The radii of the inner disks shown in \citet{Francis_2020} are mainly distributed in the range  $4$--$5$\,au, which is consistent with the radius of the WL~17 system ($<$5\,au). 
The fluxes for the inner disks reported in \citet{Francis_2020} are also comparable to that for the WL~17 system. 
Thus, the size and flux (or mass) for the WL~17 system are in good agreement with those for the other inner disks in \citet{Francis_2020}.
The flux ratio for the outer ring to the inner disk of the WL~17 system is also consistent with those in \citet{Francis_2020}. 
The WL~17 system and other systems (or objects) are different with regard to their ratio of inner disk radius to outer ring radius.  
The ring radii for the systems in \citet{Francis_2020} range from 31 to 258 au, with an average of 102.2 au and a median of 84 au. 
On the other hand, the outer ring radius (or outermost radius of the ring) is 21\,au for the WL~17 system, 
smaller than those for the systems reported in \citet{Francis_2020}, suggesting that  WL~17 may be younger than the other objects (for relevant discussion, see \S\ref{sec:dispersal}). 
Note that the sample from \citet{Francis_2020} is biased to larger and brighter protoplanetary disks, and the median disk size in Ophiuchus is smaller than that in their sample. 
Thus, further samples may be necessary to discuss the youth of the disks.

\begin{figure*}[t]
    \centering
    \includegraphics[width=0.95\linewidth]{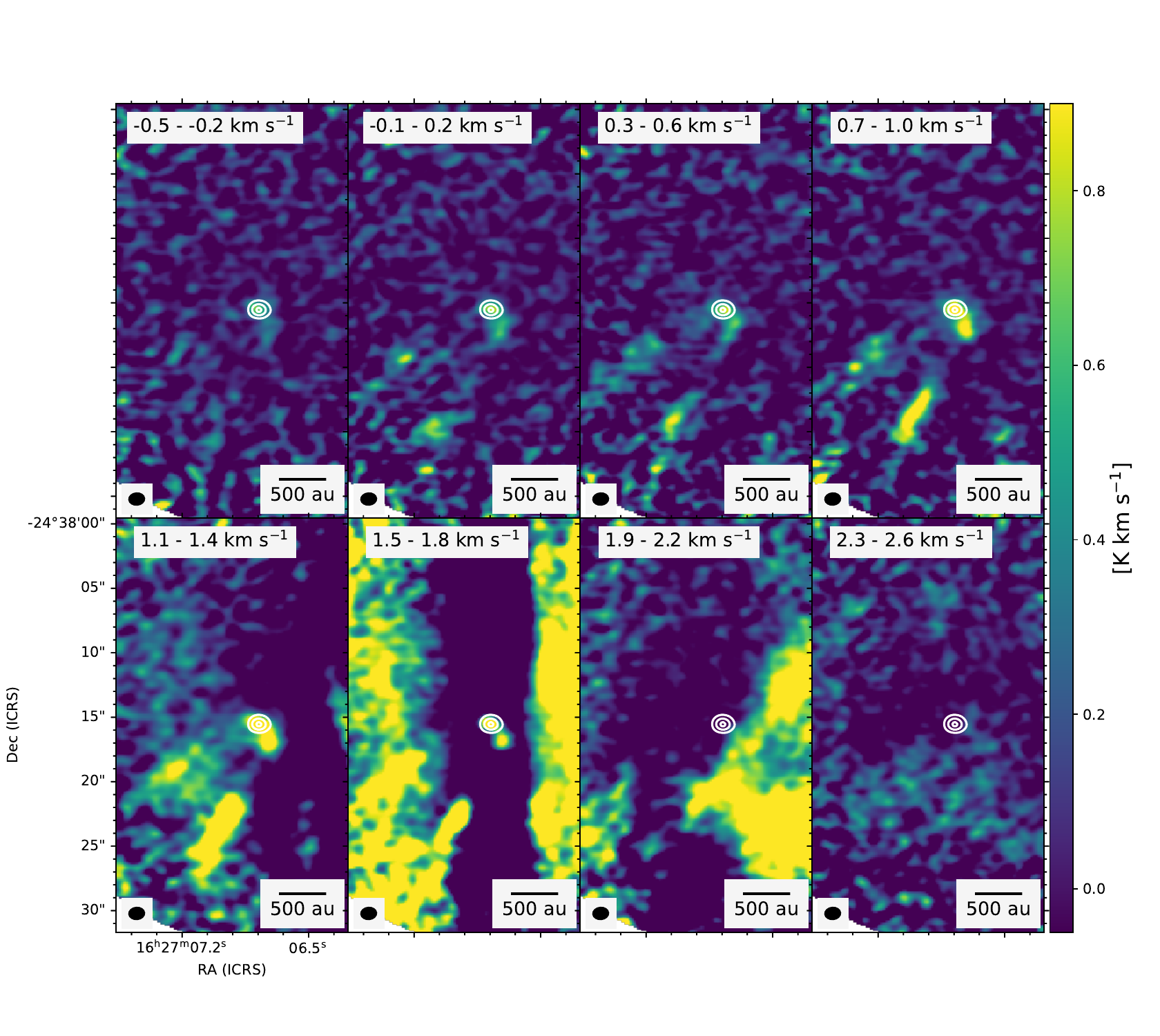}
    \caption{Velocity-channel maps of $^{12}$CO\,($J$ = 2--1) blue-shifted emission toward WL~17.
    The system velocity is $v_{\rm sys}=$ 4\,km\,s$^{-1}$ \citep{van_der_Marel_2013}.
    The integrated velocity ranges are shown at the top of each panel.
    The white contours show the 1.3\,mm continuum emission with 30, 70, and 110 $\sigma$ ($1\sigma = 3.3 \times 10^{-4}$\,Jy\,beam$^{-1}$).
    The black ellipse in the lower left corner of each panel represents the synthesized beam size of 1\farcs2 $\times$ 0\farcs9 (164\,au $\times 123$\,au) with a PA of $-$85.5$^{\circ}$.
}

    \label{fig:blue}
\end{figure*}

\begin{figure*}[t]
    \centering
   \includegraphics[width=0.95\linewidth]{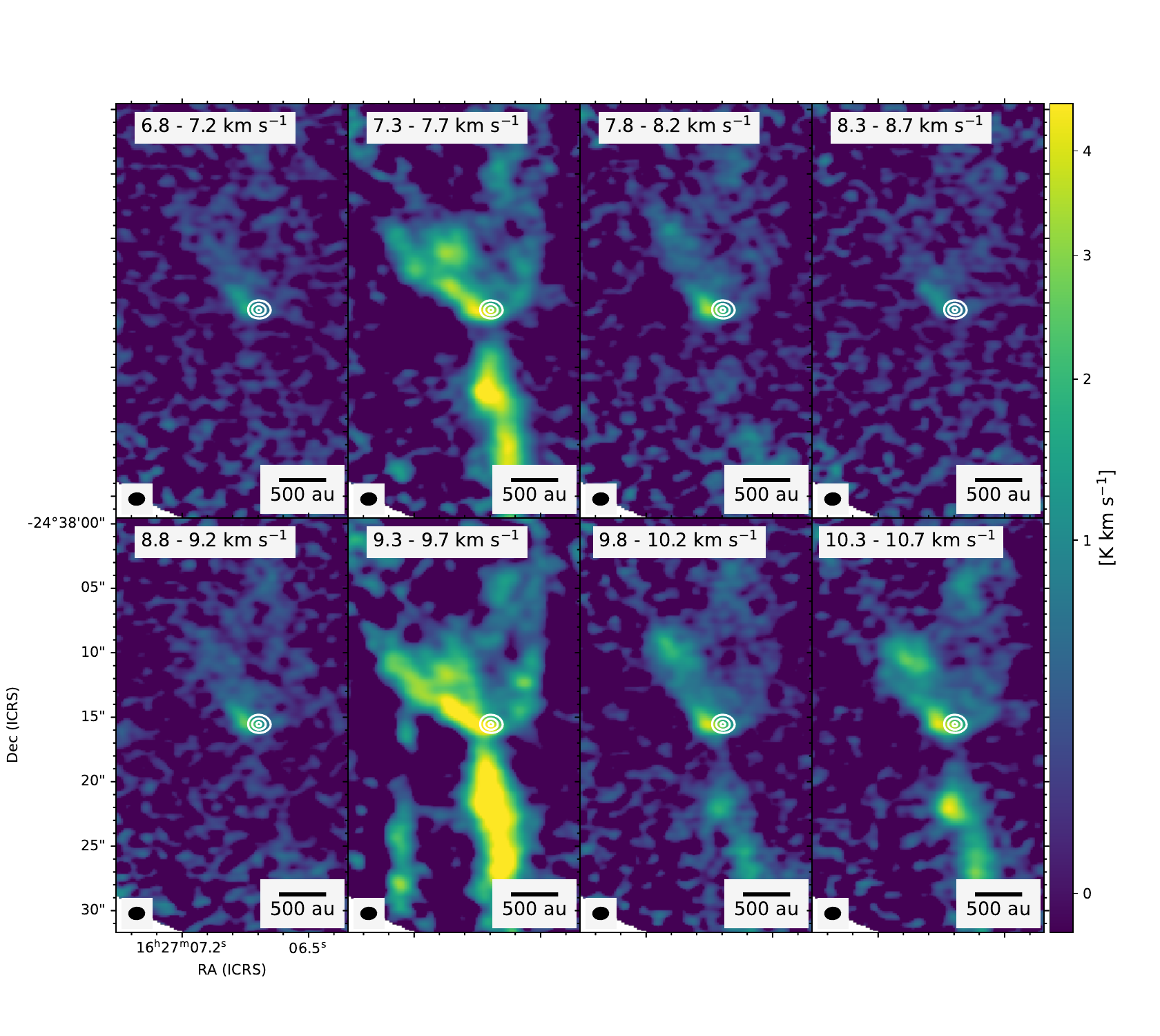}
    \caption{As Figure~\ref{fig:blue}, but for the $^{12}$CO red-shifted emission.}
    \label{fig:red}
\end{figure*}

\subsection{$^{12}$CO Line Emission and Blue- and Red-shifted Outflows}\label{outflow}
\begin{figure*}[ht]
    \centering
    \includegraphics[width=0.85\linewidth]{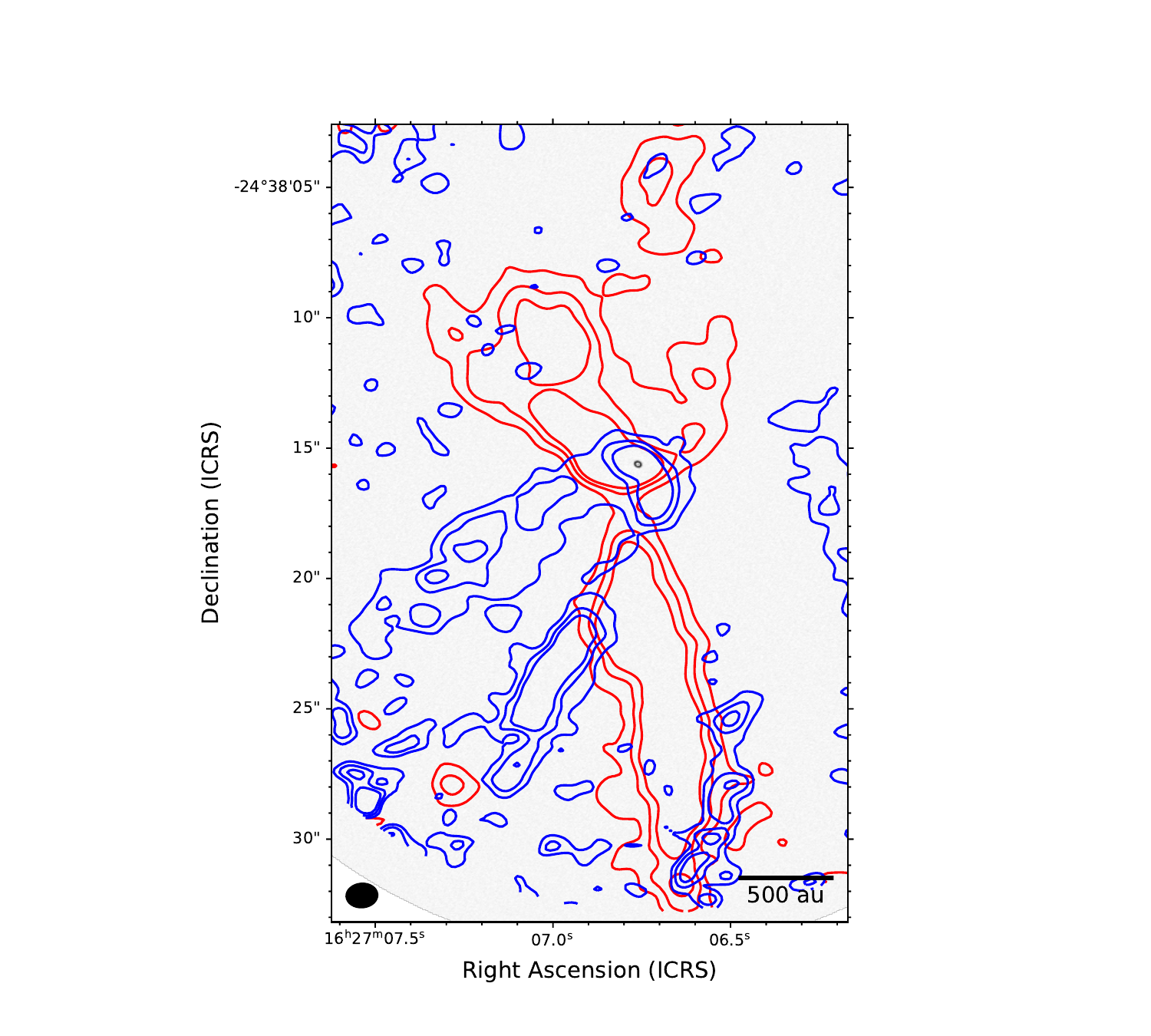}
    \caption{Moment 0 maps of blue- and red-shifted outflows with high-resolution continuum image.
    The integrated velocity ranges of the blue- and red-shifted emissions are 0.3--1.3 and 6.9--8.5\,km s$\mathbf{^{-1}}$, respectively.
    The blue contours show the $^{12}$CO emission for the blue-shifted outflow with 0.4, 1.0, 1.6\,K\,km\,s$^{-1}$.
    The red contours show the $^{12}$CO emission for the red-shifted outflow with 1.5, 3.5, 5.5\,K\,km\,s$^{-1}$.
    The black ellipse in the lower corner is the synthesized beam size (as in Fig.~\ref{fig:blue}).
    }
    \label{fig:contmom}
\end{figure*}

Figures~\ref{fig:blue} and \ref{fig:red} show the channel maps of the $^{12}$CO ($J$ = 2--1) emission. 
Figure~\ref{fig:contmom} represents the moment 0 maps of the $^{12}$CO ($J$ = 2--1) emission.
In Figures \ref{fig:blue}--\ref{fig:contmom}, we can confirm a cavity-like structure expected to be produced by a protostellar outflow.
We also determined the outflow velocity using the position velocity (PV) diagrams (for details, see Fig.~\ref{fig:pvmap}).
We call the structure shown in Figure~\ref{fig:blue} the blue-shifted outflow and that in  Figure~\ref{fig:red} the red-shifted outflow.
The blue-shifted outflow extends southeastward toward the central object (black contours) in Figure~\ref{fig:blue}.
On the other hand,  we can confirm that the red-shifted outflow cavity extends in the north direction (Fig.~\ref{fig:red}). 
In addition to the cavity-like structure, we can also see a $^{12}$CO emission extending south toward the central object in Figure~\ref{fig:red}. 
Although we cannot clearly explain the emission in the south direction, it may be a secondary outflow (for details, see \S\ref{sec:discussion}). 
The bases of the blue-shifted and red-shifted outflows are clearly superimposed on the white contour (dust continuum emission) in Figures~\ref{fig:blue} and \ref{fig:red}.
Thus, we could confirm that both the blue- and red-shifted outflows (or outflow cavities) are associated with the disk (or outer ring and inner disk) observed in the 1.3\,mm continuum emission (Fig.~\ref{fig:cont}).

\begin{figure*}[ht]
    \centering
\end{figure*}

Figure~\ref{fig:pvmap} (a) and (c) also show the integrated intensity (or moment 0) maps of the blue-shifted (left) and red-shifted (right) $^{12}$CO outflows.
Both the blue-shifted and red-shifted components show collimated (or narrow) flows enclosed by a cavity-like structure (Figure~\ref{fig:contmom}). 
As described above, the outflow directions (red-shifted and blue-shifted components) expected from the cavity-like structure are not exactly aligned each other. 
The outflow direction is in the southeast for the blue-shifted component, while it is in the northeast direction for the red-shifted component.
Different direction outflows around a single embedded protostar can be seen in recent three-dimensional simulations \citep[e.g.,][]{Matsumoto_2017} and observations \citep{Okoda_2021,Sato_2023}. 
When an inhomogeneous infalling envelope encloses the disk, the propagation directions of the outflows differ, as described in \citet{Matsumoto_2017}. 
In addition, more than one pair (blue-shifted and red-shifted)  of outflows can appear in different directions when the circumstellar disk is enclosed by an inhomogeneous infalling envelope \citep{Hirano_2020, Machida2020C}.

\subsection{Outflow Physical Quantities} \label{sec:quantities}
A large-scale outflow has been observed around WL~17, as described in \S\ref{sec:intro}. 
Using the $^{12}$CO ($J$ = 3--2) emission obtained from the James Clerk Maxwell Telescope, \citet{van_der_Marel_2013} found a strong outflow that seems to be associated with the WL~17 system, and they estimated the outflow properties. 
The blue-shifted component has a length of $r_{\rm blue, lobe}=7380$\,au, a mass of $M_{\rm blue, lobe}=2.0 \times 10 ^{-5}\,M_{\odot}$, and a maximum outflow velocity of $v_{\rm blue, max} =5.0\,\kms$.
The red-shifted outflow component has $r_{\rm red, lobe}=4500$\,au, $M_{\rm red, lobe}=4.0 \times 10 ^{-4}\,M_{\odot}$, and $v_{\rm red, max} =5.3\,\kms$.
The averaged dynamical timescale of the outflow is $t_{\rm dyn}\simeq5.5 \times\,10^3$\,yr.
The total mass ejection rate is $\dot{M}_{\rm out} \sim 10^{-7}\, M_{\odot}\, {\rm yr}^{-1}$.
The spatial resolution was not sufficient to identify the driving source of the outflow in \citet{van_der_Marel_2013}, and hence it is difficult to clearly state that the WL~17 system drives an outflow. 

\begin{figure*}[t]
    \centering
    \includegraphics[width=0.9\linewidth]{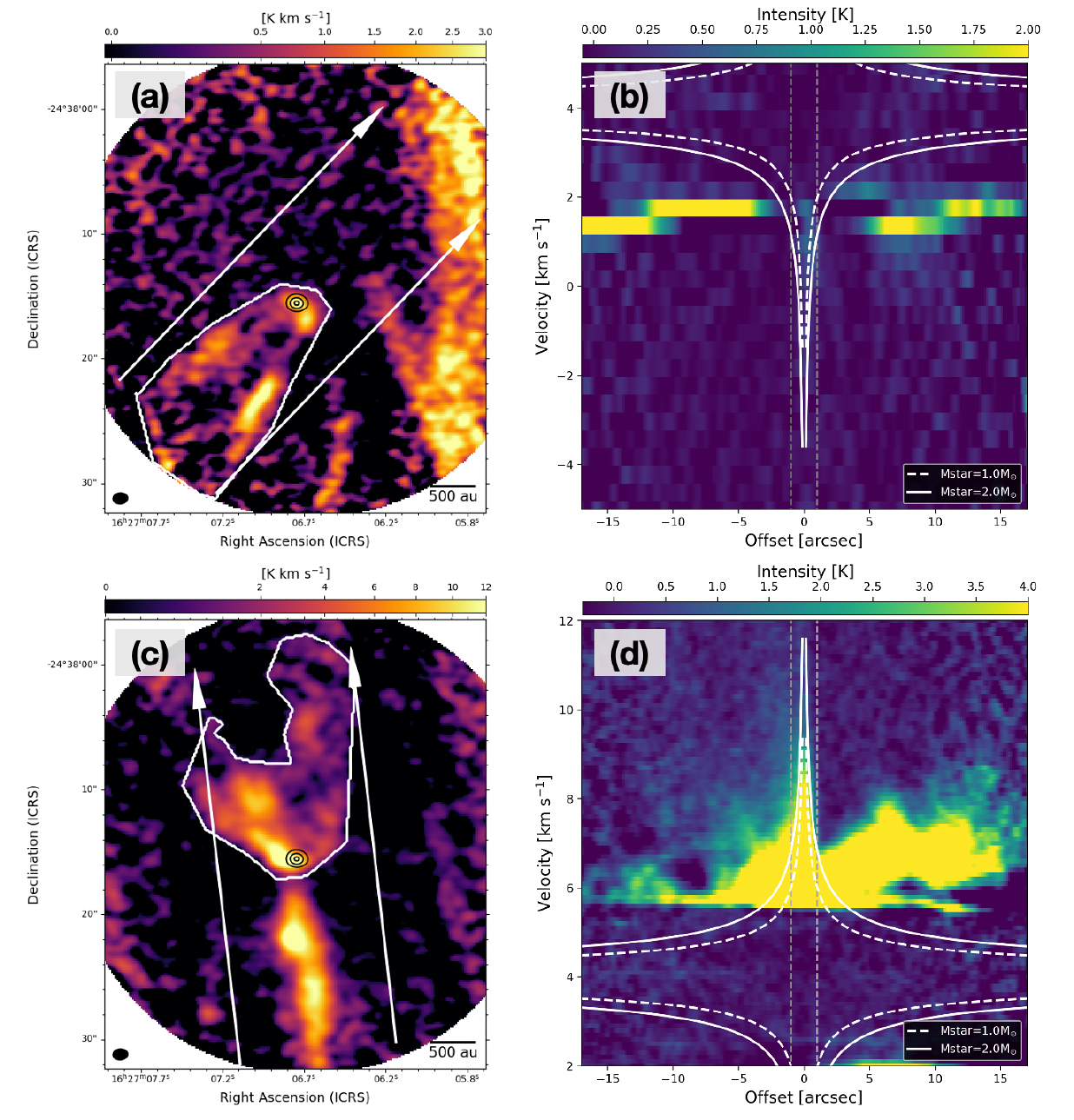}
    \caption{Spatial structures and velocity distributions of $^{12}$CO emission around WL~17.
    (a) Moment 0 map of $^{12}$CO blue-shifted emission. The black contours show the 1.3 mm continuum emission. 
    The black ellipse in the lower left corner is the synthesized beam size.
    The white arrows are the direction of the PV diagram shown in panel (b).
    (b) Position velocity diagram along the direction from southeast to northwest of the blue-shifted outflow. The direction and area used to make the PV diagram are shown in panel (a).
    The white dashed and solid lines represent the Keplerian rotation for stellar masses of 1 and 2\,$M_{\odot}$, respectively.
    The area of the protoplanetary disk (or ring--gap structure) is plotted by the vertical gray broken lines (the offsets of $\pm$1\,arcesc).
    (c) As for panel (a), but for $^{12}$CO red-shifted emission. 
    The white arrows are the direction of the PV diagram shown in panel (d).
    (d) As for panel (b), but for the red-shifted outflow. The direction and area used to make the PV diagram are shown in panel (c).
    The area enclosed by the white contours (panels a and c), which includes outflows with intensities higher than 0.5\,K\,km\,s$^{-1}$, represents the region used for calculating the outflow mass.
    }
    \label{fig:pvmap}
\end{figure*}

In this study, we could confidently confirm the base of the outflow with high spatial resolution observations. 
The outflow originates from the ring--gap structure around WL~17.
The blue-shifted (Fig.~\ref{fig:pvmap}a) and red-shifted (Fig.~\ref{fig:pvmap}c) outflows detected in the $^{12}$CO emission are shown in the left panels of Figure~\ref{fig:pvmap}, in which the areas surrounded by white lines  indicate outflow areas with intensities exceeding 0.5\,K\,km\,s$^{-1}$.
We estimated the physical properties of the outflows around WL~17 in the same manner as in \citet{van_der_Marel_2013}.

The lengths of the blue-shifted ($r_{\rm blue, out}$) and red-shifted ($r_{\rm red, out}$) outflows are almost the same $r_{\rm blue, out} \simeq  r_{\rm red, out} \simeq r_{\rm out}$ and are $r_{\rm out} \sim$\,2.4$\times$10$^{3}$\,au in size.
Note that we expect that the outflow extends further from the field of view in Figures~\ref{fig:blue} and \ref{fig:red}, because the outflow lengths estimated in our study are several times smaller  than those in \citet{van_der_Marel_2013}.
Thus, we cannot estimate the actual length of the outflow with these figures due to the limited field of view.  
It is likely that Figures~\ref{fig:blue} and \ref{fig:red} show a recent (or latest) mass ejection event. 
In the following, we estimate the physical quantities of the recent mass ejection event with  $r_{\rm out}$.

We estimate the outflow dynamical timescale $t_{\rm dyn}$ using the outflow length $r_{\rm out}$ and the maximum outflow velocity $v_{\rm max}$.  
We determined the outflow maximum velocity $v_{\rm max}$ with the system velocity $v_{\rm sys}=$ 4\,km\,s$^{-1}$ \citep{van_der_Marel_2013}.
To investigate the outflow (maximum) velocity, we generated PV diagrams along both the blue- and red-shifted outflows (Fig.~\ref{fig:pvmap}{\it b} and {\it d}), from the regions bounded by the two arrows in Figure~\ref{fig:pvmap}{\it b} and {\it d}.

The PV diagrams show that the high-velocity components are located around the dust continuum emission near the center. 
The region where the dust continuum emissions are detected is bounded by the gray broken lines in the PV diagrams. 
We also plotted the Keplerian velocity profile assuming protostellar masses of 1 and 2\,$M_{\odot}$ on the PV diagrams. 
The velocities within the gray broken line can be attributed to Keplerian rotation, assuming a protostellar mass of $1$--$2\,\msun$.
We used a protostellar mass of $1\msun$ for calculating the physical quantities for the WL~17 system below.

Although the velocities just outside  the dust continuum emission seem to significantly exceed the Keplerian velocity of 2\,$M_{\odot}$, it is difficult to identify the origin of the high-velocity components near the dust continuum emission.
The high-velocity emissions near the center could be composed of both Keplerian motion and high-velocity outflow (or jet).  
Thus, we excluded the high-velocity emissions near  the region  bounded by the gray broken lines when estimating the outflow velocity.

We can also see high-velocity components far from the dust continuum emission (or outside the region bounded by the gray broken lines). 
In both the red- and blue-shifted outflows, we could detect emissions of $\vert v - v_{\rm sys} \vert \gtrsim$\,2--5\,$\kms$ about $>1000$\,au distant from the emission peak for the dust continuum (Fig.~\ref{fig:pvmap}). 
With the PV diagrams, we determined the maximum velocity of the blue-shifted ($v_{\rm blue,max}$) and red-shifted ($v_{\rm red,max}$) components as  $v_{\rm blue,max}=$2.0$\,\kms$ and $v_{\rm red,max}=$3.5$\,\kms$, respectively.
Then, we adopted the velocity for each blue- and red-shifted outflow to calculate the mass outflow rate.
Thus, the outflow rate estimated in this study could be smaller than the actual outflow rate (for details, see below).  
The dynamical timescale $t_{\rm dyn}$ can be calculated as
\begin{equation}
t_{\rm dyn} = \frac{r_{\rm out}}{v_{\rm max}} \times \tan{(i_{\rm out})},
\end{equation}
where $i_{\rm out}$ is the inclination angle of the outflow ($= i_{\rm ring}$) and the dynamical timescales of the red-shifted $t_{\rm dyn,red}$ and blue-shifted $t_{\rm dyn,blue}$ outflows are estimated using the maximum velocity of the blue-shifted ($v_{\rm blue,max}$) and red-shifted ($v_{\rm red,max}$) components, respectively.
The outflow length, maximum velocity, and dynamical timescale are summarized in Table~\ref{table:analysis}.
The average of the dynamical timescale $t_{\rm dyn}$ between the blue-shifted and red-shifted outflows $ t_{\rm dyn}= (t_{\rm dyn,blue}+t_{\rm dyn, red})/2 $ is $t_{\rm dyn}=$\,3.0\,$\times 10^3$\,yr, which is about two times shorter than that derived in \citet{van_der_Marel_2013}. 
Thus, it is expected that the outflow shown in Figures~\ref{fig:blue} and \ref{fig:red} and \citet{van_der_Marel_2013} was ejected in the recent period $\lesssim$ 10$^3$--10$^4$\,yr. 

\begin{center}
\begin{longtable*}[ht]{cccccc}
\caption{Outflow physical quantities}\\
\hline
Outflow & $r_{\rm out}$ & $v_{\rm max}$ & $t_{\rm dyn}$ & $M_{\rm out}$ & $\dot{M}_{\rm out}$ \\
& ($10^{3}$\,au) & ($\kms$) & ($10^{3}$\,yr) & ($10^{-4}$\,$M_{\odot}$) & ($10^{-7}\,M_{\odot}$\,yr$^{-1}$)\\
\hline 
\hline
Blue lobe & 2.4 & 2.0 & 3.8 & 1.4 & 0.4 \\
Red lobe & 2.5 & 3.5 & 2.2 & 7.2 & 3.2 \\
Total & - & - & - & 8.7 & 3.6 \\
\endhead
\hline
\endfoot
\hline
\end{longtable*}
\label{table:analysis}
\end{center}

Next, we estimate the outflow mass $M_{\rm out}$ as  $M_{\rm out} = \mu m_{\rm H} AN_{\rm ave}$, where $\mu  $, $m_{\rm H}$, $A$, and $N_{\rm ave}$ are the mean molecular weight of H$_2$ ($\mu=2.8$), the hydrogen atom mass, the outflow area, and the average column density of H$_2$ in the area, respectively. 
We convert the $^{12}$CO integrated intensity $W_{\rm{CO}}$ to the H$_2$ column density $N_{\rm{ave}}$ using the $X_{\rm{CO}}$ factor, where $N_{\rm{ave}} = X_{\rm{CO}} \times W_{\rm{CO}}$. A $X_{\rm{CO}}$ factor of $X_{\rm{CO}} = 2.0 \times 10^{20}\,\rm{cm^{-2}\,K^{-1}\,km^{-1}\,s}$ is adopted \citep{Balatto_2013}.
The total mass of the blue-shifted and red-shifted outflows is $M_{\rm out,total}=$8.7$\times$10$^{-4}\, M_{\odot}$ (see also Table~\ref{table:analysis}). 
The outflow mass derived in this  study is comparable to that in \citet{van_der_Marel_2013}.

The protostellar mass of the WL~17 system can be estimated to be 1--2$M_\odot$ from the Keplerian velocity (for details, see  Fig.~\ref{fig:pvmap}).
Thus, the WL~17 system does not contain very low-mass objects, such as a proto-brown dwarf.
Note that very low-mass objects are considered to have low immensities ($< 0.1 L_{\odot}$) even in the Class 0 stage.
Assuming a constant mass loss rate of the outflows, we estimated the total mass loss rate (sum of the red and blue shifted outflows) of 3.6$\times$10$^{-7}$\,$M_{\odot}$ yr$^{-1}$, as described in Table~\ref{table:analysis}.
This mass loss rate is several times larger than that in \citet{van_der_Marel_2013}.  
The mass outflow rate $\dot{M}_{\rm out} \simeq $\,3.6$\times$10$^{-7}$\,$M_{\odot}$ yr$^{-1}$ estimated in our analysis is smaller than  that for typical Class 0 objects \citep{Nagy_2020,Feddersen_2020}.
Note that the mass outflow rates for Class 0 objects are the largest among objects with different evolutionary stages (Class 0, I, Flat, and II objects) and they decrease as the protostellar systems evolve \citep[e.g.,][]{Bontemps_1996}. 
We do not intend to claim that WL~17 is a very young, Class 0 object, 
but rather that it should be considered to be in the (later) accretion phase because the outflow physical parameters are similar to those for Class I (or Flat) objects. 
As described in \citet{Sheehan_2017}, the bolometric luminosity of WL~17 is estimated to be $L_{\rm bol}=0.6$\,$L_{\odot}$, which is comparable to that of Class I, Flat, and II objects \citep{Watson_2016}.

\section{Discussion} \label{sec:discussion}
\subsection{Remaining Mass of Core and Envelope  around WL~17}\label{sec:envelopemass}
We need to confirm the amount of mass in the infalling envelope to constrain the timescales of disk growth and ring formation, as described in \S\ref{sec:growth}. 
Thus, in this subsection, we discuss the remaining mass of the core and envelope around WL~17.

\begin{figure*}[t]
    \centering
    \includegraphics[width=\linewidth]{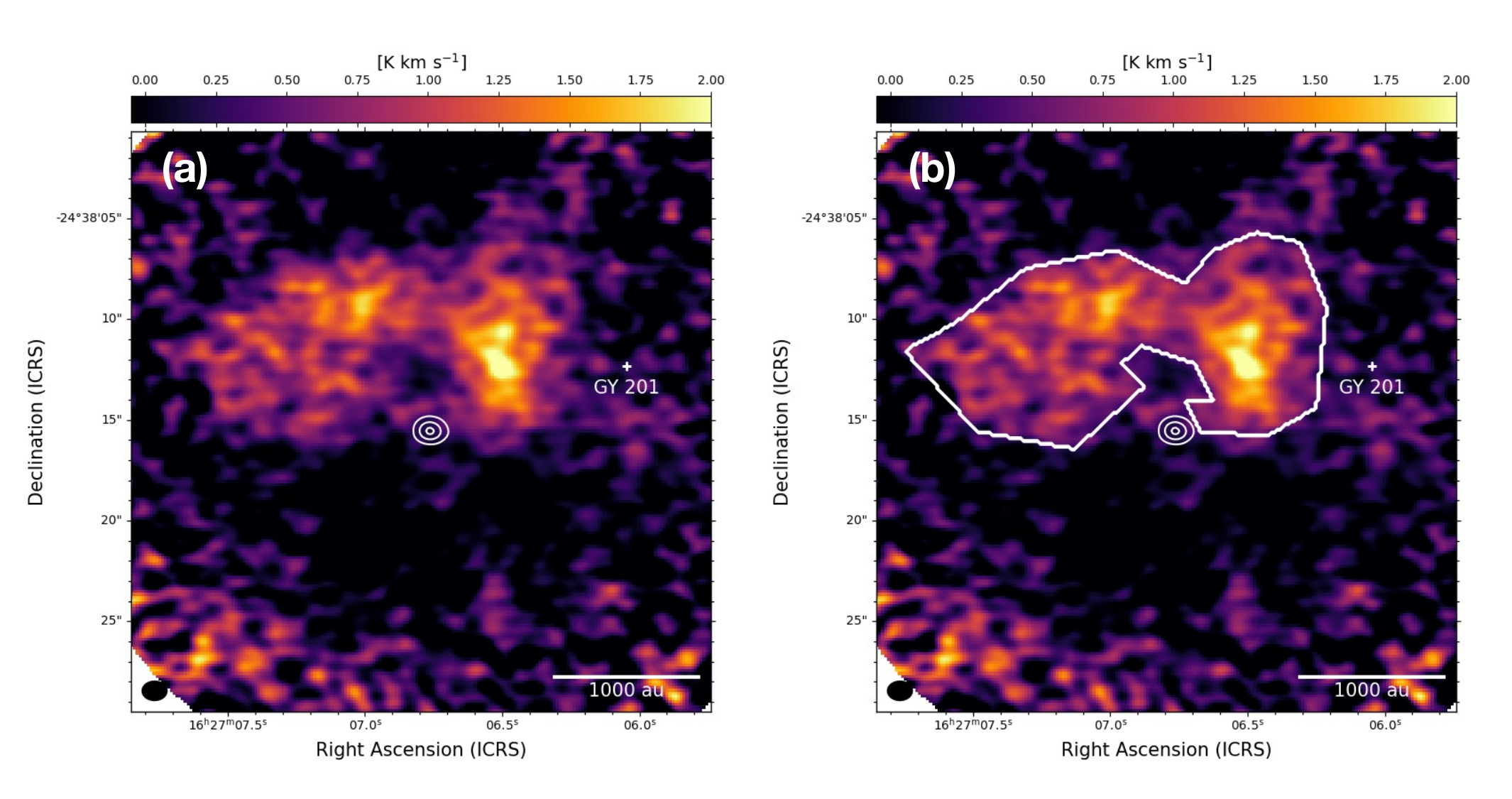}
    \caption{(Left) C$^{18}$O\,($J=2$--1) moment 0 map toward WL~17.
    The integrated velocity range is 4.0--5.4\,km\,s$^{-1}$.
    The black ellipse in the lower left corner represents the synthesized beam size of 1\farcs2 $\times$ 1\farcs0 (164\,au $\times$ 137\,au) with a PA of $-$89.4$^{\circ}$. 
    The white cross denotes the position of GY~201.
    The white contours show the 1.3\,mm continuum emission (same as Fig.~\ref{fig:blue}).
    (Right) As for the left panel, but the white line delineates the region with integrated intensity greater than 0.3\,K\,km\,s$^{-1}$. This region is used to estimate the mass of C$^{18}$O gas.
    }
    \label{fig:envelope}
\end{figure*}

As described in \S\ref{outflow}, the outflow is associated with  the WL~17 system. 
Considering that the outflow is powered by the accreting matter, it is expected that there is remaining infalling or envelope mass in the vicinity of the WL~17 system.
To investigate the envelope around WL~17, Figure~\ref{fig:envelope} shows the C$^{18}$O emission that can trace the high-density gas. 
Figure~\ref{fig:envelope} shows that the distribution of C$^{18}$O emission is asymmetric and seen primarily to the north of the object, rather than enclosing it.
The figure also indicates the existence of high-density gas in the region $\lesssim 1000$\,au around the central object WL~17.  
In addition, although the C$^{18}$O emission can be detected at the peak position of the 1.3 mm dust continuum emission, the emission is not very strong. 
The C$^{18}$O emission in the region within $\sim100$\,au from the central object is also not strong. 
Thus, Figure~\ref{fig:envelope} indicates that envelope gas still remains around the WL~17 object, while the gas distribution is not homogeneous. 
In the last decade, ALMA observations revealed a protostellar system with inhomogeneous envelope structures \citep{Pineda_2022} where the rotationally supported disk is detached from the surrounding dense material \citep[e.g.,][]{Tokuda_2017}.
Although the origin of such complex systems is still debated \citep[e.g.,][]{Matsumoto_2015,Tokuda_2018,Pineda_2022}, numerical simulations by \cite{Matsumoto_2017} and \cite{Kuffmeier_2017} demonstrated that the inhomogeneous distribution of the envelope gas could induce a time-variable accretion and mass ejection.

Assuming a protostellar mass of $M_*=1\,M_{\odot}$, the freefall velocity $v_{\rm ff}$ of the gas distributed about $r\sim1000$\,au from the central object can be estimated to be $v_{\rm ff}=(2GM_*/r)^{1/2} \sim 1.3\,\kms$. 
Thus, the envelope gas seen in Figure~\ref{fig:envelope} could fall within $r/v_{\rm ff}\sim5000$\,yr. 
Therefore, mass accretion would last for at least a further $\sim10^3$--$10^4$\,yr which is shorter than the lifetime of Class 0/I objects \citep{Enoch_2009}. 
Note that we estimated the required time using the location of WL~17 and C$^{18}$O emission projected on the plane of the sky. 

\begin{figure*}[t]
    \centering
    \includegraphics[width=\linewidth]{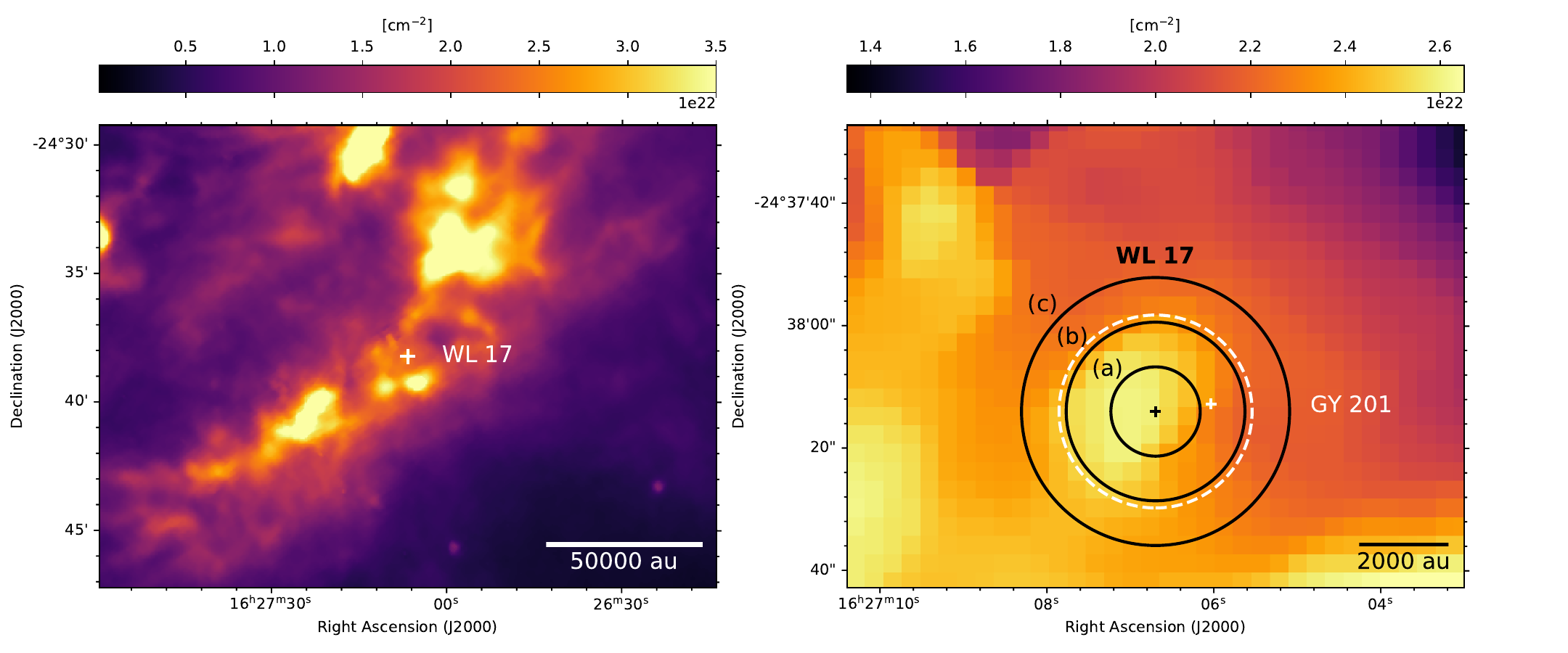}
    \caption{Herschel column density maps of the L1688 region in the $\rho$ Ophiuchus molecular cloud at the resolution of 18$"$ \citep{Ladjelate_2020}.
    In the left panel, the position of WL~17 in the L1688 region is indicated by the white cross.
    The right panel is an enlarged view of the left panel around WL~17.
    The black and white crosses indicate the positions of WL~17 and a nearby young stellar object (GY201), respectively.
    The dashed circle shows the observation field for our ALMA data.
    The radii of 1000, 2000, and 3000\,au are represented by solid black circles, with WL~17 as the origin.
    The enclosed masses within 1000, 2000, and 3000\,au of WL~17 are described in Table~\ref{table:envelope}.} 
    \label{fig:herschel}
\end{figure*}

Figure~\ref{fig:herschel} shows the column density map derived from the Herschel Gould Belt Survey data \citep{Andre_2010} with a resolution of 18$"$ \citep{Ladjelate_2020}.
The figure indicates that WL~17 is located within a large-scale filamentary structure with a length of $\sim1$\,pc and width of 0.1\,pc (Fig.~\ref{fig:herschel} left). 
Within the filament, there are some clumps and young stellar objects (YSOs).
We can see a small high-density clump with a size of $\sim 2000$--$3000$\,au associated with WL~17 (Fig.~\ref{fig:herschel} right).   
Although the column density outside the core where WL~17 is embedded is relatively low, that of the WL~17 core is high.  
As seen in Figure~\ref{fig:herschel}(b), GY 201 is located close to WL~17. 
However, the dense gas traced by C$^{18}$O (Fig.~\ref{fig:envelope}) seems to be associated with WL~17. 
Thus, although not all the matter shown in the right panel of Figure~\ref{fig:herschel} would accrete onto the WL~17 system, a non-negligible amount of the matter is associated with the WL~17 system.  
Therefore, based on the outflow detection, the C$^{18}$O distribution around WL~17 and the Herschel data, we expected that WL~17 is in the accretion phase of star formation.

We measured the remaining (or envelope) mass around WL~17.
We estimated the mass of molecular hydrogen (H$_2$) using the C$^{18}$O column density.  
First, we need to determine the optical depth $\tau_\nu$, which can be  calculated using the following equation \citep{Frerking_1982}: 

\begin{equation}
    T_B=\frac{h\nu}{k} \left(\frac{1}{e^{h\nu/{kT_{\rm ex}}}-1}-\frac{1}{e^{h\nu/{2.7k}}-1}\right)\left(1-e^{-\tau_{\nu}}\right),
    \label{eq:tb}
\end{equation}
where $h$ is the Planck constant, $k$ is the Boltzmann constant, $\nu$ represents the spectral frequency ($\nu$=220\,GHz), and
$T_{\rm ex}$ denotes the excitation temperature, which is assumed to be 
10 and 20\,K.
$T_{\rm B}$ corresponds to the brightness temperature obtained from the observed data in C$^{18}$O, which is around 1.1\,K.
We performed a Gaussian fit of the spectra in the area of Figure~\ref{fig:envelope} (b) and used the FWHM as the velocity width $\Delta v\,(\sim 0.8$\,km\,s$^{-1}$).
With the derived optical depth $\tau_\nu$ and the velocity width $\Delta v$, we can estimate  the C$^{18}$O column density $N_{\rm C^{18}O}$ as 

\begin{equation}
    N_{\rm C^{18}O}=\frac{3k}{4\pi^3\nu\mu_0^2} \frac{T_{\rm ex} \exp({\frac{J}{2}\frac{h\nu}{kT_{\rm ex}}})}{1-\exp({-\frac{h\nu}{kT_{\rm ex}}})}\tau_\nu \Delta v,
    \label{eq:C18Ocd}
\end{equation}
where we assume that the line spectrum has the Gaussian velocity dispersion, the $\mu_0$ is the dipole moment and $J$ is the energy level \citep{Frerking_1982}.
The C$^{18}$O column density $N_{\rm C^{18}O}$ is then converted to the averaged H$_2$ column density $N_{\rm H_2}$ using the relation $N_{\rm C^{18}O}=1.7\times10^{-7}\times N_{\rm H_2}$ \citep{Frerking_1982}.
Finally, we calculate the converted H$_2$ mass as follows:
\begin{equation}
    M_{{\rm H}_2}=\mu m_p N_{{\rm H}_2} \Delta S,
    \label{eq:H2mass}
\end{equation}
where $\mu$ (= 2.8) is the mean molecular weight of H$_2$ \citep{Kauffmann_2008}, $m_p$ is the proton mass and $\Delta S$ is the area enclosed by the thick white line in Figure~\ref{fig:envelope}(b).
The mass of molecular hydrogen estimated from C$^{18}$O is $4\times10^{-3}\,\msun$.
We can also estimate the mass of molecular hydrogen using the Herschel column density maps shown in Figure~\ref{fig:herschel}(b) with equation~(\ref{eq:H2mass}).
We calculated the mass within each black circle in Figure~\ref{fig:herschel}(b), which have radii of 1000, 2000, and 3000\,au, respectively.  
Because the WL~17 system is located in the filament, the H$_2$ profile cut in the southwest direction across WL~17 has a sharp decline and a flat distribution at $\sim$1.5$\times$10$^{22}$\,cm$^{-2}$.
The flat (column) density is widely distributed in the region 0.1\,pc away from WL~17, where 0.1\,pc corresponds to the typical dense core size and filament width \cite[e.g.,][]{Onishi_2002,Arzounamian_2011,Tokuda_2019}. Therefore, we treated the column density of 1.5$\times$10$^{22}$\,cm$^{-2}$ as the  background level.
The enclosed mass within these radii are listed in Table~\ref{table:envelope}.

The total mass of molecular hydrogen in the vicinity of WL~17 is as small as $M_{\rm H_2}=3.6\times10^{-3}$--$0.11\,\msun$(Table~\ref{table:envelope}). 
Note that, within $r\lesssim 1000$\,au, the molecular hydrogen mass estimated from the C$^{18}$O emission seems to be significantly smaller than that from Herschel column density. 
This difference can be partly attributed to missing flux in the ALMA observation \citep[e.g.,][]{Tokuda_2018}. 
Furthermore, the inhomogeneous distribution of CO emission (Fig.~\ref{fig:envelope}) can also cause the difference in the estimated molecular hydrogen masses from the ALMA and Herschel observations.  
However, a substantial amount of mass ($M_{\rm H_2}=0.1\msun$) remains in the envelope within 2000--3000\,au of WL~17 (Table~\ref{table:envelope}). 
Although a Class II source GY~201 may be embedded in the same condensed gas region around WL~17, Figures~\ref{fig:envelope} and \ref{fig:herschel} suggest that the dense gas is primary  associated with the WL~17 system. 
Thus, mass accretion onto WL~17 is expected to continue in future evolutionary stages.
The remaining mass surrounding the WL~17 system ($\sim0.1\msun$) is relatively small compared to the  protostellar mass of WL~17 ($\sim1\msun$). 
Therefore, we expect that the WL~17 system is in the late stage of the Class I phase, indicating the end of the mass accretion stage.

\begin{center}
\begin{longtable*}[ht]{ccccc}
\caption{Envelope mass around WL~17}\\
\hline
   & Calculation area & $N_{{\rm H}_2}$ & $\Delta S$ & $M_{{\rm H}_2}$\\
   & & ($\times$10$^{22}$\,cm$^{-2}$) & ($\times$10$^7$\,au$^2$) & ($M_\odot$) \\
\hline
\hline
C$^{18}$O ($T_{\rm ex}$=10\,K) & White-line area in Figure~\ref{fig:envelope}\,(b) & 0.4 & 0.3 & 4.4$\times$10$^{-3}$\\
C$^{18}$O ($T_{\rm ex}$=20\,K) & White-line area in Figure~\ref{fig:envelope}\,(b) & 0.3 & 0.3 & 3.6$\times$10$^{-3}$\\
H$_2$     & Black circle (a) in Fig.~\ref{fig:herschel}, $R < 1000$\,au & 1.0$^\ast$ & 0.3 & 1.0$\times$10$^{-2}$\\
H$_2$     & Black circle (b) in Fig.~\ref{fig:herschel}, $R < 2000$\,au & 1.0$^\ast$ & 1.3 & 5.5$\times$10$^{-2}$\\
H$_2$     & Black circle (c) in Fig.~\ref{fig:herschel}, $R < 3000$\,au & 0.9$^\ast$ & 2.8 & 1.1$\times$10$^{-1}$\\
\endhead
\hline
\endfoot
\multicolumn{5}{l}{{\footnotesize $\ast$. The averaged H$_2$ column density $N_{\rm H_2}$ has been background subtracted (the background level$\sim$1.5$\times$10$^{22}$\,cm$^{-2}$).}}\\
\endlastfoot
\hline
\end{longtable*}
\label{table:envelope}
\end{center}

\subsection{Disk Growth Timescale}
\label{sec:growth}
In this subsection, we discuss the disk growth timescale during the mass accretion phase in terms of star formation.
Ring and gap structures are usually observed in Class II objects. 
For Class II objects, the envelope is already depleted and the mass accretion and outflow rate are as low as $\ll 10^{-7}$\,$M_{\odot}$\,yr$^{-1}$ \citep{Hartmann_1998}.  
Thus, it is considered that mass transfer from the envelope onto the disk has already been completed for Class II objects. 
Assuming a disk mass of $M_{\rm disk, II}=0.01$\,$M_{\odot}$ and a mass accretion rate onto the disk from the envelope of $M_{\rm acc, II}=10^{-8}$\,$M_{\odot}$\,yr$^{-1}$ for Class II objects, the disk growth timescale is $t_{\rm growth, II}\sim M_{\rm disk, II}/\dot{M}_{\rm acc, II}\sim10^{6}$\,yr. 
In addition, the dissipation timescale of protoplanetary disks is considered to be $\sim10^6$--$10^7$\,yr \citep{Fedele_2010}. 
Thus, we can consider that the ring and gap structures form within $\lesssim 10^6$--$10^7$\,yr around Class II objects. 

On the other hand, when the mass accretion rate onto the disk is high, as in Class 0 and I objects, the disk grows in a short time.  
Since the ring and gap structure could be related to dust growth and planet formation, it is crucial to specify when dust growth and planet formation begin. 
Therefore, it may be valuable to properly identify the evolutionary stage of WL~17 in order to constrain the timescale for ring and gap formation, as described in \S\ref{sec:intro}.
Although various mechanisms have been proposed for the formation of ring and gap structures, we focus on the dynamic formation of ring and gap structures in an early or accretion stage of star formation.

During the mass accretion phase, the disk mass continues to increase due to the continuous mass supply from the infalling envelope. 
However, the disk cannot possess a large amount of mass because the disk becomes gravitationally unstable, and the disk material, composed of gas and dust, rapidly accretes onto the central protostar. 
Thus, the disk material is replaced on a short timescale (or disk growth timescale).
Even if the dust ring is formed, it will fall onto the central protostar with the gas component when the rapid mass accretion induced by gravitational instability occurs.
Thus, even if the ring and gap form in the mass accretion phase, they will only survive only within the disk growth timescale, during which all the disk material (gas and dust) should be replaced \citep[for details, see the review by][]{Tsukamoto_2023ASPC}.

When the mass accretion rate from the envelope to the disk is high, the time required for ring and gap formation becomes short.
Note that the ring and gap must form before the pre-existent disk material is replaced by the newly accreting matter as described above.
The mass accretion rate $\dot{M}_{\rm acc}$ is assumed to be 10 times the mass loss rate  $\dot{M}_{\rm acc} \simeq \varepsilon \dot{M}_{\rm out}$, where $\varepsilon=10$ is adopted \citep{Carbit_2009,Konigl_2000}. 
Thus, with the mass outflow rate derived in \S\ref{sec:quantities}, the mass accretion rate is estimated to be $\dot{M}_{\rm acc}\simeq10^{-6}\,M_{\odot} \,{\rm yr}^{-1}$.
Although the mass accretion rate derived from the outflow rate is 3.6$\times10^{-6}$\,$M_{\odot}$\,yr$^{-1}$ with $\varepsilon =10$, we conservatively use $\dot{M}_{\rm acc}\simeq10^{-6}\,M_{\odot} \,{\rm yr}^{-1}$ in the following order of magnitude analysis.
Assuming a disk mass of $M_{\rm disk}=0.01\,M_{\odot}$, the disk growth timescale is as short as $(M_{\rm disk}/\dot{M}_{\rm acc}) \sim 10^4$\,yr, indicating that the accreting matter replenishes the disk within $\sim 10^4$\,yr.
In this case, rapid formation of the ring and gap structure (outer ring and inner disk) is required because the gas and dust within the disk are replaced by the accreting matter roughly every $\sim10^4$\,yr.  
Part of the disk gas is ejected by the outflow and the remainder falls onto the central star. 
Note that if the original disk matter remains within the disk without falling,  the disk mass continues to increase as mass accretion proceeds.
Then, gravitational instability occurs, which rapidly enhances the mass accretion rate onto the central star from the disk, as described above.   
As a result, the disk mass during the main mass accretion phase (or Class 0/I stage) is adjusted so as to balance the accreting matter from the envelope and the matter falling onto the central star \citep{Tomida_2017}. 
In other words, fresh gas and dust continue to feed the disk, while those of the previous generation are pushed into the central star. 
Thus, the gas and dust cannot stay in the disk for a long time with a high mass accretion rate. 
Therefore, when a ring and gap structure appears in the accretion stage (or Class 0/I stage), we can severely constrain the timescale for dust growth and planet formation.

\subsection{Possible Formation Scenarios for Ring Gap Stricture around WL~17}
\label{sec:scenarios}
In this study, for the first time, we discovered an inner disk around WL~17 enclosed by the ring--gap structure. 
In addition, we found that an outflow is associated with the WL~17 system based on the $^{12}$CO emission, and the mass ejection rate due to the outflow is comparable to those for Class I objects. 
We also confirmed that the infalling envelope remains around the WL~17 system.
Furthermore, we did not detect any strong emission or noticeable structure within the gap. 
With these new findings, we constrain the evolutionary stage and formation process of the ring--gap structure.
In the following, we introduce and discuss four possible scenarios for the production of the ring--gap structure around WL~17: embedded binary system, unseen protoplanet, dispersal by disk wind, and dust growth front.

\subsubsection{Embedded Binary System}
First, we discuss the scenario for ring--gap formation by a binary system.
When a binary system exists in a high-density region (or is embedded within an inner disk), the gravitational torque due to the binary orbital motion produces a non-axisymmetric structure that can transport the angular momentum outward.
As a result,  a ring--gap-like structure can form in a region outside the binary system, as shown in observations \citep[e.g.,][]{Takakuwa_2017} and numerical simulations \citep[e.g.,][]{Matsumoto_2019}. 
For this scenario, a non-axisymmetric structure should appear outside the binary system to enhance the mass accretion and produce a gap \citep{Machida_2009,Matsumoto_2019}.  
Thus, a spiral-like structure corresponding to the channel flow to the binary system should be detected within the ring or gap. 
In addition, it is expected that the emission (or density distribution) of the ring-like structure is not uniform because the binary motion disturbs the circumbinary disk \citep{Saiki_2020}. 
The ring enclosing the binary system corresponds to the circumbinary disk in this scenario. 
Circumbinary disks tend to appear in the embedded phase of star (or binary) formation because the mass of the circumbinary disk can be supplied from the infalling envelope. 

In \citet{Sheehan_2017}, the region inside the ring or gap cannot be sufficiently resolved. 
Thus, with their observation, it is difficult to entirely reject the existence of a spiral structure created by the binary system in the gap region. 
On the other hand,  we could detect neither a clear spiral nor non-axisymmetric structure in the gap. 
In other words, we could not confirm any sign of a binary system in the high-spatial-resolution image shown in Figure~\ref{fig:cont}. 
However, we could not resolve the inner disk with a sufficiently high spatial resolution. 
In addition, if a binary system is embedded in the inner disk, it is possible for the secondary outflow described in \S\ref{outflow} to be driven by a binary companion \citep{Kuruwita_2017,Saiki_2020}. 
Thus, although  we cannot completely disprove the existence of a very close binary embedded in the inner disk, the embedded binary scenario is not plausible for explaining the ring--gap structure around WL~17.

\subsubsection{Unseen Protoplanet}
The most straightforward scenario to explain the ring--gap structure is a protoplanet in the gap. 
When a protoplanet orbits in a disk,  a gap forms around the planet orbit \citep[][]{Kanagawa_2015,Kanagawa_2016, 2017MNRAS.469.1932D}. 
The detection of a circumplanetary disk in a gap or hole is strong evidence of the existence of a protoplanet. 
Recently, a circumplanetary disk was detected around PDS 70 \citep{Keppler_2018,Bensity_2021} 
and two protoplanets (or circumplanetary disks) could be confirmed. 
The structure of PDS 70 is similar to that of WL~17. 
The PDS 70 system has a ring--gap structure in which an inner disk has also been detected. 
Although no circumplanetary disk could be detected in the gap around WL~17, it might be seen in future observations with higher sensitivity and spatial resolution.
In addition, it is possible for a small planet with a mass of several Earth masses to produce a ring--gap structure when the disk viscosity is considerably small \citep{Muto_2010,Perez2019}.
For example, \citet{Crida2006} showed that the gap opening mass for an inviscid disk can be described as 
\begin{equation}
\frac{M_{\rm p}}{M_*} \gtrsim 1.6\times10^{-4} \left( \frac{H/r}{0.05}
\right),
\end{equation}
where $M_{\rm p}$, $M_*$, $H$ and $r$ are the planet mass, protostellar mass, disk scale height, and radius, respectively \citep[see also][]{Muto_2010}. 
Thus, a small planet can create a gap depending on the disk parameters. 
Therefore, even if we do not detect a circumplanetary disk in the gap with future observations, we cannot reject the scenario in which a 
 (small-sized) protoplanet is embedded within the gap.

The difference between PDS 70 and WL~17 is the evolutionary stage or age of the central object. 
The age of PDS 70 has been estimated to be $\sim5.4$\,Myr, thus in the Class II or III stage \citep{Muller_2018}. 
On the other hand, WL~17 is considered to be in the accretion phase (Class I of the Flat stage). 
The age of WL~17 is estimated to be $<0.5$\,Myr \citep{Evans_2009}, and thus it is much younger than PDS 70. 
Thus, forming a planet in such a short timescale is challenging. 
We need further observations to verify the invisible protoplanet scenario.

\subsubsection{Dispersal by Disk Wind}
\label{sec:dispersal}
To reproduce the ring--hole structure of WL~17 shown in \citet{Sheehan_2017}, \citet{Takahashi_2018} proposed a scenario of disk dispersal by the disk wind. 
Based on a viscous (or $\alpha$) disk model, they considered the mass loss from the disk by an magneto-rotational instability (MRI) wind \citep{Suzuki_2009}.
Their study determined the dust size so that a Stokes number of ${\rm St}=0.1$ is realized at each radius, indicating that the dust size differs at different radii. 
The gas in the disk is dispersed from the inner disk region by the disk wind because the dispersal timescale depends on the Keplerian timescale. 
In other words, since the Keplerian timescale becomes short as the distance from the center decreases, the dispersal timescale is shorter for a smaller radius than for a larger radius. 
Thus, the hole is created from the inside out in the disk.   
The gas density or pressure is enhanced at the dispersal front. 
Then, a pressure bump (or pressure maximum) appears.
Since the dust accumulates around the pressure bump, a ring--hole structure forms.

They investigated disk evolution from the prestellar phase (i.e., prestellar cloud core with a size of $\sim0.1$\,pc) and showed that the ring--hole structure forms in a short timescale of $\lesssim 10^5$\,yr. 
Thus, the ring formation timescale is compatible with our results. 
However, no inner (dust) disk appears in their fiducial model because the dust falls onto the central star before the gas is dispersed by the wind.
Note that, for their fiducial model, they assumed that the dust drift timescale is comparable to or shorter than the disk dispersal timescale in the inner disk.
Thus, the existence of the inner disk in our analysis is not compatible with their fiducial model. 

\citet{Takahashi_2018} reproduced an inner (dust) disk in other models.
They showed that a ring--gap structure (or inner disk) forms when the dust drift timescale is longer than the disk wind (or dispersal) timescale in the inner disk region.
In this case, the disk wind removes the gas from the inner region. 
Thus, dust radial drift does not occur due to a deficit of gas because the dust grains do not feel gas drag in the inner region. 
Therefore, the inner disk is composed only of dust without a gas component. 
This model seems to be compatible with our new findings shown in \S\ref{sec:results}. 
However, the MRI disk wind is slower than the CO outflow shown in Figures~\ref {fig:blue} and \ref{fig:red} \citep{Suzuki_2009}.
Instead of the MRI disk wind, we may have to consider a magnetocentrifugal wind \citep{Bai_2017,Tabone_2020} because the velocity of the wind needs to exceed the escape velocity of the system.  
In any case, we can verify the disk wind scenario to confirm whether abundant gas exists in the inner disk region.
This scenario requires depletion of gas in the inner disk region. 
The gas depletion in the central region may explain the low bolometric luminosity of WL~17.

\subsubsection{Dust Growth Front}
\label{sec:front}
While dust growth was not considered in \citet{Takahashi_2018}, it is possible to explain the structure around WL~17 in terms of dust growth.  
\citet{Ohashi_2021} showed that the ring structure around WL~17 can be explained by a dust growth front within which the dust radial drift becomes effective and the dust grains are depleted. 
The dust growth timescale is proportional to the Keplerian timescale.
Thus, assuming the same sized dust grains at the epoch of disk formation, the dust grows from the inside out. 
\citet{Ohashi_2021} also showed that the dust growth front is observed as a ring in their synthetic observation. 
In addition, they showed that a ring corresponding to the dust growth front appears within $\sim10^3$--$10^4$\,yr, which is comparable to the disk growth timescale estimated in this study (\S\ref{sec:envelopemass}).
Since dust growth is a key to producing the ring-like structure, we can verify this scenario by measuring the spectral index of dust continuum emissions as described in \citet{Ohashi_2021}.
Recently, \citet{Ilseing_2023} presented the ALMA Band 3 (3\,mm) and 7 (0.87\,mm) observations of the WL~17 system.
They identified an off-center hole and an asymmetric ring in the Band 7 observation, which differs from the disk substructure observed in Band 3 \citep{Gulick_2021}.
In addition, they showed an asymmetric map of the spectral index with a low mean value of $\alpha$=2.28$\pm$0.02, indicating dust growth and segregation on the disk.

The growth front scenario seems to be suitable for explaining the ring--hole structure for WL~17 shown in \citet{Sheehan_2017}, while it cannot explain the inner dust disk found in this study. 
\citet{Ohashi_2021} did not consider collisional fragmentation between dust grains.  
If small dust grains are produced by collisional fragmentation, the infalling dust grains could be coupled with the gas again in the inner disk region where the gas density is high and the dust grains are prevented from such a rapid radial drift and survived.
Mass accretion onto the disk and mass ejection from the disk were also ignored in \citet{Ohashi_2021}, while our analysis indicated a  mass ejection from the WL~17 system. 
Thus, there are some inconsistencies between \citet{Ohashi_2021} and our analysis.
However, the dust growth front is a promising scenario to explain the structure around WL~17 because the ring-like structure appears in the dust thermal emission.

\subsection{Ring Structure Formation through Star Formation}
\label{sec:starformation}
As described in \S\ref{sec:envelopemass}, we estimated the disk growth timescale to be $t_{\rm grow} \sim 10^4$\,yr. 
We also assume that the formation timescale of the outer ring $t_{\rm ring}$ is comparable to the disk growth timescale $t_{\rm ring}\sim t_{\rm grow} \sim 10^4$\,yr,  because the matter in the disk is replaced by the accreting matter every $\sim10^4$\,yr.
Most Class II objects are considered to form a ring structure within  $\sim 10^6$--$10^7$\,yr.
Therefore, the WL~17 system may severely constrain the ring-gap formation timescale because the ring formation timescale is much shorter than the timescales for Class II objects and protoplanetary disks.
It has been considered that WL~17 is an important object for understanding when the dust structure (ring and gap) and planets form because the age of WL~17 has been estimated to be as short as $\lesssim 10^5$--$10^6$\,yr. 
If the ring--hole or ring--gap structure can be related to dust growth or planet formation, we can conclude that planet formation begins at an earlier stage than previously thought. 
Since dust growth and planet formation have usually been considered in an isolated disk detached from the (infalling) envelope or the remnant of the nascent cloud core, we may need to revisit the planet formation theory constructed since the 1980s. 

We simply explain the WL~17 system in terms of star formation because WL~17 is in the accretion phase of star formation. 
We confirmed that the protostellar outflow is driven by the WL~17 system.
The outflow mass ejection rate is  as high as $\gtrsim 10^{-7}\,M_{\odot}$\,yr$^{-1}$. 
If we ignore mass accretion onto the disk, a disk mass of $\sim10^{-2}\,M_{\odot}$ could be dispersed by the protostellar outflow within $\lesssim10^5$\,yr. 
Thus, mass accretion onto the disk should occur to preserve the disk. 
To confirm the infalling envelope that supplies the mass to the circumstellar disk, we detailed the dense gas envelope enclosing the WL~17 system. 
As described in \S\ref{sec:envelopemass}, the disk wind theory indicates that the mass accretion rate is about ten times the mass ejection rate, i.e., $M_{\rm acc}\gtrsim 10^{-6}\,M_{\odot}$, which corresponds well to the mass accretion rate in the main accretion phase \citep{Larson_2003}.
In this case, the disk growth timescale is estimated to be as short as $\sim10^4$\,yr, during which the gas and dust that exist in the disk at an epoch are all replaced by newly accreting matter.   
If this is the case, the ring--gap structure can form within $\sim 10^4$--$10^5$\,yr. 

Recently, \citet{Koga_2022} showed that dust grains accreted on a disk can orbit for at least $\sim10^3$\,yr without inward radial drift, because the dust grains in the outer disk region receive angular momentum from the gas and dust grains in the inner region during the main accretion phase.  
Since the disk surface density is high around Class 0 and I protostars, the dust growth timescale becomes  as short as $30$\,yr (3000\,yr) for $1\mu$m (1\,cm)-sized dust grains \citep{Koga_2022}.
Thus, the dust growth timescale is comparable to or shorter than the disk growth timescale  $\sim10^4$\,yr \citep[see also][]{Ohashi_2021}.
Therefore, dust growth or planet formation may occur in the circumstellar disk around Class 0 and I protostars during the accretion phase.

If dust growth or planet formation occurs in the disk around a Class 0 and I protostar, they  may finally fall onto the central star within the disk growth timescale $\sim10^4$\,yr. 
A steady accretion onto the disk is usually considered in the accretion phase. 
However,  the gas distribution of the infalling envelope is inhomogeneous and the gas density in the proximity of WL~17 is relatively low, as shown in Figure~\ref{fig:envelope}. 
It should be noted that a secondary outflow or misaligned outflow can appear in such an inhomogeneous envelope \citep{Hirano_2020,Machida2020C}. 
Mass accretion onto the disk from the infalling envelope may transiently stop with the inhomogeneous envelope.  
When the gas supply from the envelope onto the disk is very small, mass accretion from the disk onto the central star should also be small. 
The low luminosity of the central protostar can be explained by non-steady accretion, as seen in \citet{Vorobyov_2006} and \citet{Machida_2011Z}.  
As a result, it may be possible to observe ring--gap structures formed in the past $\sim10^4$\,yr. 
The disk becomes massive as the mass accretion rate onto the disk increases.
Then, gravitational instability occurs and a significant part of the disk gas should fall onto the central region accompanying the dust ring. 
Thus, we may be observing a temporary structure of the dust distribution in the accretion phase.
If this is the case, the dust ring, growing dust grains, and planets may only remain in the disk for a short time. 
However, if dust growth occurs on such a short timescale ($\sim10^4$\,yr) during the accretion phase, the growing dust grains and the resulting planets formed just after the  accretion phase could remain for a long time in the protoplanetary disk.

\subsection{Uncertainty in Timescale Estimation}
\label{sec:uncertainty}
The aim of this study is to constrain the ring formation or
disk mass growth timescale around WL~17, an object in the accretion stage.
As described in \S\ref{sec:growth} and \S\ref{sec:starformation}, we estimated the timescale for ring or disk mass growth as $t_{\rm grow}\sim10^4$\,yr using the following equation:  
\begin{equation}
t_{\rm grow} = \frac{M_{\rm disk}}{\dot{M}_{\rm acc}}
 = \frac{M_{\rm disk}}{\varepsilon \dot{M}_{\rm out}}
 = \frac{M_{\rm disk} r_{\rm out}}{\varepsilon M_{\rm out} v_{\rm out}},
\label{eq:timescale}
\end{equation}
where $M_{\rm disk}$ is the disk mass around WL~17, which was estimated from the dust continuum emission. 
The mass accretion rate onto the WL~17 system $\dot{M}_{\rm acc}$ is represented by 
\begin{equation}
\dot{M}_{\rm acc}= \varepsilon \dot{M}_{\rm out}, 
\end{equation}
where $\varepsilon$ (= 10) is the parameter that relates the mass accretion rate $\dot{M}_{\rm acc}$ to the mass outflow  rate $\dot{M}_{\rm out}$  \citep{Carbit_2009,Konigl_2000}. 
Recent numerical simulations have shown that the parameter is in the range $3\lesssim \varepsilon \lesssim 10$ \citep{Machida_2012,Matsushita_2017}.  
Thus, the uncertainty in the growth timescale caused by the parameter $\varepsilon$  is a factor of $\sim3$. 
The mass outflow rate  $\dot{M}_{\rm out}$ is described as 
\begin{equation}
\dot{M}_{\rm out} = \frac{M_{\rm out}}{t_{\rm dyn}},
\end{equation}
where $M_{\rm out}$ is the outflow mass and the dynamical timescale $t_{\rm dyn}$ is calculated as 
\begin{equation}
t_{\rm dyn} = \frac{r_{\rm out}}{v_{\rm out}} \times \tan{(i_{\rm out})},
\end{equation}
where $r_{\rm out}$ is the outflow length and $v_{\rm out}$ is the outflow velocity.
\footnote{Determining the inclination angle is important for estimating $t_{\rm dyn}$. However, theoretical studies have shown that it is difficult to accurately determine the inclination angle when the outflow axis is not aligned \citep{Hirano_2020,Machida2020C}.
We adopted a smaller value for the outflow velocity $v_{\rm out}$, resulting in a decrease in the mass outflow rate $\dot{M}_{\rm out}$.
Thus, it is considered that we did not significantly overestimate the mass outflow rate (or mass accretion rate) as long as the inclination angle is not extremely small.}
The outflow quantities ($M_{\rm out}$, $r_{\rm out}$ and $v_{\rm out}$) were estimated from the $^{12}$CO emission. 
We used four observation quantities, the disk mass $M_{\rm disk}$, outflow mass $M_{\rm out}$, outflow length $r_{\rm out}$ and outflow velocity $v_{\rm out}$, to estimate the disk growth timescale $t_{\rm grow}$. 
In this subsection, we discuss the uncertainty in these observational quantities and the resultant timescale. 

The disk mass derived from the dust thermal emission was estimated to be $M_{\rm disk}=M_{\rm gas,ring} \simeq 0.01\,\msun$ (\S\ref{sec:cont}). 
This may be an underestimation because we only estimated the emission from the outer ring and inner disk, 
although for sources other than these, the remaining emission is not strong. 
However, the disk mass could be overestimated if the dust-to-gas mass ratio is larger than 1/100, which is assumed to estimate the gas component of the disk.
In addition, the disk mass  $\sim0.01\,\msun$ for the WL~17 system is comparable to that for Class I objects and larger than that for Class II objects \citep[e.g.,][]{Yen_2015, Tychoniec_2020}. 
If the disk mass is larger than $0.1\msun$, the growth timescale would exceed $10^6$\,yr. 
In such a case, we cannot severely constrain the ring formation timescale because the accretion stage should end within $10^6$\,yr. 
On the other hand, our claim about ring formation in the accretion stage holds  as long as the disk mass is smaller than $0.1\,\msun$.

The existence of the outflow is evidence of mass accretion because the outflow is powered by the release of gravitational energy from the accreting matter. 
We determined the mass and velocity of the outflow from  $^{12}$CO observations (Figs.~\ref{fig:blue}-\ref{fig:pvmap}). 
The outflow mass may be underestimated because the $^{12}$CO emission may not trace the entire mass of the outflow.
We conservatively adopted an outflow velocity of $2$ and $3.5\,\kms$ to calculate the mass ejection rate (Table~\ref{table:analysis}). 
Thus, we will likely underestimate both the mass and velocity of the outflow. 
A larger outflow mass and  higher outflow velocity shortens the ring growth timescale as described in equation~(\ref{eq:timescale}). 
Thus, we expect that the actual ring formation timescale is shorter than our estimated value of $t_{\rm grow} \sim10^4$\,yr.  
In other words, the disk growth or ring formation timescale of $t_{\rm grow}\sim10^4$\,yr gives an upper limit for the WL~17 system. 
Note that we investigated the recent mass ejection event within $r_{\rm out}$.
Thus, we can arbitrarily define or determine the outflow length $r_{\rm out}$ for estimating the outflow mass and velocity, which implies that we do not need to consider the uncertainty of the outflow length.

Finally, we discuss the uncertainty in the masses of the protostar $M_*$ and envelope $M_{\rm env}$.
We roughly estimated the protostellar mass as $M_*=1\msun$ from the PV diagram (Fig.~\ref{fig:pvmap}).
However, it is difficult to separate the Keplerian motion from the high-velocity jet, as described in \S\ref{sec:quantities}.  
Thus, in this study, the protostellar mass is uncertain and cannot be determined accurately. 
The envelope mass is estimated to be $\sim10^{-3}$--$10^{-2}\,\msun$ within the region $r<1000$\,au and $\sim0.1\msun$ within the region $r<3000$\,au. 
The mass ratio of the protostar to the envelope determines the evolutionary stage of the WL~17 system. 
Since the protostellar masses appears to dominate the envelope mass, we concluded that the WL~17 system is in the late accretion stage. 
However, if the envelope mass around WL~17 is depleted or not gravitationally bound, mass accretion does not occur.
In such a case, the ring--gap structure would be maintained for a long time.
On the other hand, if there is enough infalling envelope mass around WL~17, mass accretion continues. 
In such a case, the ring--gap structure would eventually disappear in a short timescale.  
Although determining the evolutionary stage of the WL system is important, it is not essential for estimating the ring growth timescale.

\section{Summary}\label{sec:summary}
We investigated the WL~17 system using dust continuum, $^{12}$CO and C$^{18}$O lines emissions taken from  ALMA archival data. 
Previous observations have shown that WL~17 is a Class I protostar associated with protostellar outflow. 
In addition, \citet{Sheehan_2017}  found a ring--hole structure around WL~17. 
The spatial resolution of the data used in this study is higher than in previous studies.
We also confirmed a ring in the range 11--21\,au from the central object based on the dust continuum emission. 
For the first time, we detected an inner disk with a radius of $<$5\,au around the central object. 
Thus, WL~17 has a ring--gap structure, not a ring hole. 
In addition, we could not detect any 1.3\,mm dust continuum emission within the gap. 
In other words, we could not find any sign of the existence of planets or a binary companion in the dust continuum emission.  

Using the $^{12}$CO emission, we confirmed that the outflow detected in a past low-resolution observation can be associated with  the WL~17 system. 
Although the outflow axis between the blue-shifted and red-shifted components is slightly misaligned, we clearly showed outflow cavity structures in both components. 
In addition, we found another component of the outflow in the red-shifted emission 
and estimated the outflow physical quantities. 
The mass ejection rate due to the outflow is as high as $\gtrsim 10^{-7}\,M_{\odot} \,{\rm yr}^{-1}$.
Thus, we can clearly state that mass ejection is currently occurring from the WL~17 system.    

From the C$^{18}$O emission and Herschel column density map, we showed that the WL~17 system is  surrounded by infalling gas.
Thus, this study also confirms that WL~17 is in the accretion  phase of star formation.
However, the distribution of the envelope gas around the WL~17 system is inhomogeneous, which may result in a non-steady mass accretion and anisotropic mass ejection. 

We estimated the mass accretion rate onto the disk, based on the mass ejection rate estimated in this study ($\dot{M}_{\rm out} \simeq 10^{-7}\,M_{\odot} \,{\rm yr}^{-1}$). 
Assuming a mass accretion rate  of $\dot{M}_{\rm acc}\sim10^{-6}\,M_{\odot} \,{\rm yr}^{-1}$ (10 times the mass ejection rate) and a disk mass of $M_{\rm disk}\sim10^{-2}\,M_{\odot}$, the disk growth timescale is estimated to be $t_{\rm grow} \sim10^4$\,yr. 
Thus, the ring--gap structure may form in $\sim10^4$\,yr. 
Since the infalling envelope remains around the WL~17 system, the ring--gap structure may be cleaned out in the next rapid mass accretion event. 
Our results indicate that a ring--gap structure can form on a short timescale of $\sim10^4$\,yr. 
The rapid formation of the ring structure is compatible with both disk-dispersal and dust growth front scenarios. 
However, neither scenario can adequately  explain the existence of the inner disk detected in this study.
Although an unseen planet can create a gap--ring structure, we could not find any sign of the existence of a planet. 
Our findings can strongly constrain the dust ring--gap and planet formation scenarios, while we need further high-sensitivity and high-spatial resolution observations to fully unveil the formation of the structure around WL~17.

\begin{acknowledgments}
The authors wish to thank Drs. Ohashi for their helpful contributions. 
This work was supported by a NAOJ ALMA Scientific Research Grant (No. 2022-22B), Grants-in-Aid for Scientific Research (KAKENHI) of the Japan Society for the Promotion of Science (JSPS; grant nos. JP17H06360, JP17K05387, JP17KK0096, JP21K13962, JP21H00049, JP21K03617, JP21H00046, and JP22J11129). 
This paper makes use of the following ALMA data: ADS/JAO. ALMA\#2019.1.00458.S and \#2019.1.01792.S. ALMA is a partnership of the ESO (representing its member states), the NSF (USA), and NINS (Japan), together with the NRC (Canada), MOST, ASIAA (Taiwan), and KASI (Republic of Korea), in cooperation with the Republic of Chile. The Joint ALMA Observatory is operated by the ESO, AUI/NRAO, and NAOJ. Data analysis was in part carried out on a common-use data analysis computer system at the Astronomy Data Center, ADC, of the National Astronomical Observatory of Japan.

\end{acknowledgments}

\bibliography{shoshi_ref}
\bibliographystyle{aasjournal}



\end{document}